\documentclass[twocolumn]{aa}
\usepackage{graphicx}
\usepackage{times}
\begin{document}
   \title{Detached binaries in the Large Magellanic Cloud}
   \subtitle{A selection of binaries suitable for distance determination\thanks{Table 3 is only available in electronic form
             at the CDS via anonymous ftp to cdsarc.u-strasbg.fr (130.79.128.5)
             or via http://cdsweb.u-strasbg.fr/cgi-bin/qcat?J/A+A/}}
   \author{G.~Michalska, A.~Pigulski}
   \offprints{G.~Michalska}
   \institute{Instytut Astronomiczny Uniwersytetu Wroc{\l}awskiego, Kopernika 11, 51-622 Wroc{\l}aw, Poland\\
              \email{michalska@astro.uni.wroc.pl, pigulski@astro.uni.wroc.pl}}
   \date{Received 9 November 2004; accepted 14 December 2004}
   \abstract{As a result of a careful selection of eclipsing binaries in the Large Magellanic Cloud
      using the OGLE-II photometric database, we present a list of 98 systems that are suitable targets
      for spectroscopic observations that would lead to the accurate determination of the distance to the LMC.
      For these systems we derive preliminary parameters combining the
      OGLE-II data with the photometry of MACHO and EROS surveys.  In the selected sample, 58 stars have
      eccentric orbits.  Among these stars we found fourteen systems showing apsidal motion.  The
      data do not cover the whole apsidal motion cycle, but follow-up observations
      will allow detailed studies of these interesting objects.
   }
   \titlerunning{Detached binaries in the LMC}
   \maketitle
\section{Introduction}
Diverse methods have been employed in the last decade to derive distances to the Magellanic
Clouds (see, e.g., Cole \cite{cole}, Harries et al.~\cite{harries}, Alves \cite{alves}).  The knowledge
of these distances is important for at least two reasons.  First, Magellanic Clouds
play a key role in the calibration of the Cepheid-based distance scale.  Second, due to the
low metallicity of the Clouds, it can be checked how different methods used for distance
determination depend on metallicity.  Unfortunately, as far as the distance moduli of the Magellanic Clouds
are concerned, different methods gave results that disagree at a level that
leaves in doubt their applicability for more distant objects.

However, the idea of measuring distances by means of eclipsing binaries seems very promising
and supposedly will soon succeed
in finding the correct answer.  The method has been known for many years and first applied to
a Magellanic Cloud star by Bell et al.~(\cite{bell91}).  Its potential usefulness has been recalled
a few years ago by Paczy\'nski (\cite{paczynski}), while historical outlook was
presented by Kruszewski \& Semeniuk (\cite{krusz}).  The method
exploits the advantages of a combination of the light curve of an eclipsing
binary with its double-lined spectroscopic orbit.
This combination provides radii, masses and ratio of surface brightnesses of the
components.  To get the distance, it is enough to calibrate surface brightness, $F_{\rm V}$,
in terms of an easily-observed colour index and account for interstellar extinction.
The calibration of $F_{\rm V}$ is usually made using nearby (that is, having accurate parallaxes)
eclipsing binaries and/or interferometric and occultation radii of stars
(Barnes et al.~\cite{barnes78}, di Benedetto \cite{dibe}, van Belle \cite{vanbe}).

In order to avoid possible systematic effects on the calibration of $F_{\rm V}$ for close binaries,
it was pointed out at the beginning that {\it detached\/} systems (hereafter DEBs) are the best for the purpose of distance
determination (e.g.~Paczy\'nski
\cite{paczynski}).  However, since alternatively to the application of the calibration
of $F_{\rm V}$, the fit of the modern model atmospheres to the UV/optical spectrum can be
utilized, the advantage of using only the DEBs is not so straightforward.  As was
argued by Wyithe \& Wilson (\cite{WW2}) and Wilson (\cite{wils04}), {\it semi-detached\/} binaries can be used for the
purpose of distance determination as well.  Unlike in the DEBs, the photometric mass
ratio can be derived, so that a single-lined spectroscopic orbit is sufficient.  In addition,
current modeling programs describe the proximity effects very well.  Moreover, the orbits of semi-detached systems
are well circularized, which is not always the case for the DEBs.  In fact, out of ten
eclipsing systems used by Harries et al.~(\cite{harries}) to get the distance to the Small
Magellanic Cloud (SMC), only three are the DEBs, while six have a semi-detached geometry.
Since each binary provides the distance, one can get the
average distance with an unprecedented accuracy using as many systems as possible.
The ten systems analyzed by Harries et al.~(\cite{harries})
led to the---so far---best estimation of the SMC distance modulus: 18.89 $\pm$ 0.04~mag with a
0.10-mag uncertainty due to systematic effects.  The authors, however, intend to derive the
distance of the SMC using over 100 eclipsing binaries.

A similar work is underway for the Large Magellanic Cloud (hereafter LMC).   So far, seven systems were used to get its
distance.  Out of them, only three, HV\,2274 (Guinan et al.~\cite{guinan}, Udalski et al.~\cite{udala},
Ribas et al.~\cite{hv2274}, Nelson et al.~\cite{nelson}, Groenewegen \& Salaris \cite{groen},
Fitzpatrick et al.~\cite{hv982}), EROS\,1044 (Ribas et al.~\cite{eros1044}) and HV\,982 (Fitzpatrick et al.~\cite{hv982},
Clausen et al.~\cite{clausen}), are DEBs.
Out of the remaining four, three, i.e.~HV\,5936 (Bell et al.~\cite{bell93}, Fitzpatrick et al.~\cite{fitz03}), HV\,2241 (Pritchard et
al.~\cite{prit98a}, Ostrov et al.~\cite{ostr01}) and HV\,2543 (Ostrov et al.~\cite{ostr00}) are semi-detached systems, whereas
Sk\,$-$67$^{\rm o}$\,105
(Ostrov \& Lapasset \cite{osla}) is even a contact binary.
As indicated above, in order to derive the distance accurately, we need to analyze many more systems.

Fortunately, microlensing surveys delivered photometric measurements for millions of stars
in the Clouds and catalogs including thousands of eclipsing binaries were published.
First, Grison et al.~(\cite{grison}) published double-band photometry for 79 eclipsing systems
in the LMC bar discovered within the EROS survey. Next,
Alcock et al.~(\cite{alcock}) classified 611 eclipsing binary stars
in the LMC from the MACHO survey. Finally, the catalogs of eclipsing binaries from the OGLE-II
survey were published: for the SMC (Udalski
et al.~\cite{udalb}, Wyrzykowski et al.~\cite{wyrz-smc}), and the LMC (Wyrzykowski et al.~\cite{wyrz-lmc}, hereafter W03).
In total, about 4500 eclipsing binaries (1914 in the SMC and 2580 in the LMC) were found in the OGLE-II
data in both Clouds.
Therefore, the first step towards the accurate distance determination to the LMC is good selection of
binaries that are suitable for this purpose.

This is the main goal of this paper.  Using the photometry obtained during the second
phase of the OGLE survey, OGLE-II, we have selected 98 DEBs brighter than 17.5~mag in $V$.
We have also analyzed their light curves by means of the Wilson-Devinney (Wilson \& Devinney \cite{wd},
Wilson \cite{wils79}, \cite{wils90})
program, combining the photometry from the OGLE-II, MACHO and EROS surveys.  The preliminary results
of this paper have been presented by Michalska \& Pigulski (\cite{our-prel}). A list of 36 DEBs
in the LMC good for distance determination was presented by W03.
A similar selection has been done for the SMC eclipsing binaries by Udalski et al.~(\cite{udalb}), Wyithe \& Wilson (\cite{WW1}),
and Graczyk (\cite{graczyk}).

The data used in this study are described in Sect.~2. Section 3 presents our selection criteria followed by
description of the new transformation of the OGLE-II differential fluxes to magnitudes (Sect.~4).
The analysis of the light curves and discussion of the parameters is described in Sect.~5.  Finally,
we discuss systems with apsidal motion (Sect.~6) and provide our conclusions (Sect.~7).

\section{The photometry}

\subsection{OGLE data}

The OGLE observations we used were carried out during the second phase of this microlensing
survey (Udalski et al.~\cite{ogle2}) with the 1.3-m Warsaw telescope at the Las Campanas Observatory, Chile. They cover
about 4.5 square degrees in the LMC bar (twenty one 14$^{\prime}$ $\times$ 57$^{\prime}$ fields).
The data span almost four years (1997--2000).  This was the primary source of our photometric
data used to select the DEBs (Sect.~3).  The other databases (MACHO and
EROS) were used as supplementary ones.

The OGLE-II photometry for the LMC fields is currently available in three different forms:
\begin{itemize}
\item[$\bullet$] Mean $BVI$ magnitudes for $\sim$7 $\times$ 10$^6$ stars (Udalski et al.~\cite{udal00}) derived by means of the
profile-fitting package {\sc DoPhot} (Schechter et al.~\cite{dophot}).
\item[$\bullet$] Time-series $I$-filter photometry of about 53\,400 variable candidates published by
\.Zebru\'n et al.~(\cite{zebr-cat}).  The photometry was obtained by means of the Difference Image Analysis
(DIA) package developed by Wo\'zniak (\cite{wozniak}) which is an implementation of the
image subtraction method of Alard \& Lupton (\cite{ism}).  The photometry is available from the
OGLE web page\footnote{http://www.sirius.astrouw.edu.pl/\~{}ogle/ogle2/dia/} in two forms: as
differential fluxes and the magnitudes transformed from these fluxes.  The transformation was
explained in detail by \.Zebru\'n et al.~(\cite{zebr-cat}).
\item[$\bullet$] $BVI$ time-series photometry for the same variable candidates obtained with {\sc DoPhot} and available
from the same web page.  Typically, about 30, 40 and
400 datapoints are available for each star in $B$, $V$ and $I$ bands, respectively.
\end{itemize}

Since the OGLE data were made in the drift-scan mode (Udalski et al.~\cite{ogle2}), for a given frame
the average epoch of observation is not the same for all stars. Following the prescription given
by \.Zebru\'n et al.~(\cite{zebr-cat}), we applied appropriate corrections.  Moreover,
the data were phased with the derived orbital period and the outliers were
rejected.

\subsection{MACHO data}

The MACHO observations were obtained in blue (440--590 nm) and red (590--780 nm)
bands, rougly coincident with Johnson $V$ and Cousins $R$,
respectively.  The telescope used was the 1.27-m Great Melbourne Telescope situated
at Mount Stromlo, Australia.  The MACHO data span almost eight years between 1992 and 2000.
The photometry is available through the MACHO web page (Allsman \& Axelrod \cite{allsman}).

The MACHO data are affected by the presence of a large number of outliers.  We therefore
rejected them in an iterative process fitting accurately the shape of the light curve in the phase
diagram.  In addition, a spurious variation of instrumental origin with a 1-yr period was removed prior to
the analysis.  Then, the heliocentric corrections were applied to the published epochs.  Since the latter
corresponded to the beginning of exposures, half the exposure time (150~s) was added to the
epochs as well.

\subsection{EROS data}

The EROS observations were made in 1991--1992 with the 0.4-m telescope at La Silla, Chile,
in two bands, $B_{\rm E}$ and $R_{\rm E}$,
having central wavelengths of 490 and 670~nm, respectively (Grison et al.~\cite{grison}).
Since the latter corresponds roughly to the MACHO red band, we used in our study
the photometry made in the $B_{\rm E}$ band only.
The epochs of EROS data were published in local time (see Ribas et al.~\cite{eros1044}), so that
a three-hour correction was added.  As for the OGLE and MACHO data, some outliers were
removed prior to analysis.

\section{Selection of objects}

\begin{figure}
\includegraphics[width=8.7cm,clip]{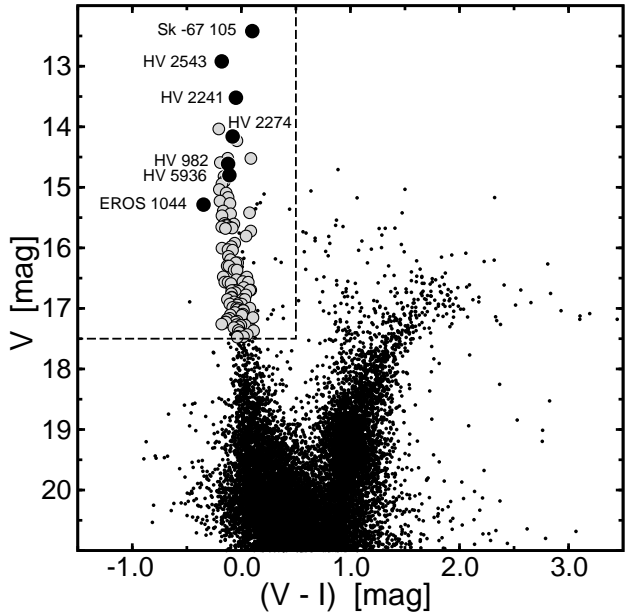}
\caption{The colour-magnitude diagram for $\sim$16\,000 stars from the OGLE-II SC$\_$1 field (small dots).
The 98 DEBs we selected as good targets for distance determination are plotted as gray filled circles.
For comparison, seven eclipsing binaries already used for the determination of the distance to the LMC
are plotted as black filled circles and labeled.  The dashed line delimits our selection box.}
\label{cmd}
\end{figure}

As we pointed out in previous sections, the OGLE-II database of candidate variables
(\.Zebru\'n et al.~\cite{zebr-cat}) was the primary source
of the analyzed data and the subject of the main selection process.  From this catalog, we
first extracted the $I$-filter photometry of those stars that were brighter than 17.5~mag in $V$ and had
$V-I <$ 0.5~mag (Fig.~\ref{cmd}).  For all these stars, the analysis of variance (AoV) periodograms of
Schwarzenberg-Czerny (\cite{alex}) were calculated. Then, the data were phased with
the period corresponding to the frequency of the maximum peak in the periodogram and examined
visually.  After this check, the star was selected for further analysis if: (i) the light curve
indicated it was an EA-type eclipsing binary, presumably a DEB, (ii) the proximity effects were small,
(iii) the scatter in the light curve was relatively low.

In total, we found 98 stars that meet our three criteria.  For all but five the MACHO
photometry is available, eleven are in the list of eclipsing binaries found within the EROS
survey.   The stars are listed and cross-identified in Table \ref{tab-list}.

Our selection was made independently of the work of W03,
who---using the same data---discovered 2580 eclipsing binaries, including 1817 of the EA type.
It was, however, not our aim to make a complete catalog of eclipsing binaries in the LMC, but to select
the brightest DEBs with the best photometry.  There are 403 EA-type eclipsing binaries in the catalogue
of W03 that fall into the selection box in Fig.~\ref{cmd}.
Thus, the stars we selected constitute about 25\% of this sample.  However, three objects from our list,
\#24, \#87, and \#91\footnote{The numbers preceded by `\#' refer to the first column of Table 3, available
only in electronic form from the CDS.}, were not included in the W03 catalogue.
\begin{figure}
\centering
\includegraphics[width=8.7cm,clip]{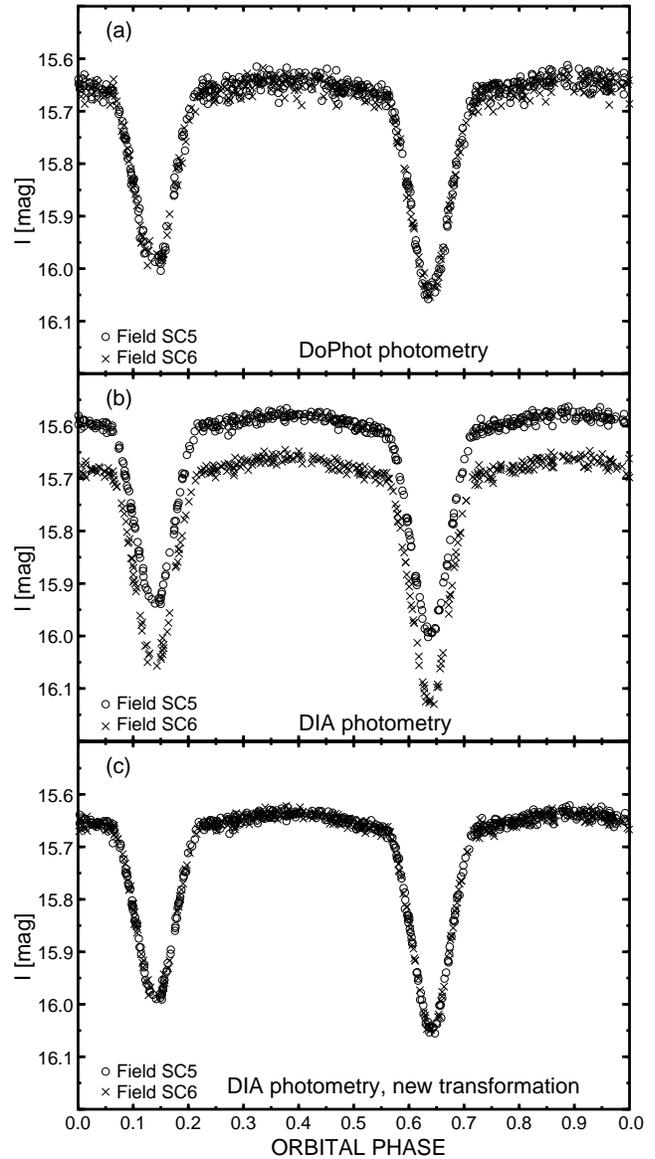}
\caption{$I$-filter light curves of the OGLE\,05223546$-$6931434 (\#17) eclipsing binary observed in two fields, LMC\_SC5
(circles) and LMC\_SC6 (crosses). (a) {\sc DoPhot} photometry, (b) DIA photometry, (c) DIA photometry with new
transformation to magnitudes and a 0.003-mag shift between zero points applied. Note the reduction of scatter
in this diagram with respect to the light curve shown in panel (a).}
\label{dia-dophot}
\end{figure}

\begin{table*}
\centering
\caption{List of 98 selected DEBs in the Large Magellanic Cloud. $P_{\rm orb}$ stands for the
orbital period.  The $VI$ photometry was derived from the OGLE-II {\sc DoPhot} data and is given
for the out-of-eclipse phase at maximum light.
In the last column `E' was given for system with non-zero eccentricity and `AM' for
systems with detectable apsidal motion.  For systems that were indicated as
good targets for distance determination by W03, the number is followed by an asterisk.  Stars are arranged according to the
decreasing $V$ magnitude.}
\label{tab-list}
\begin{tabular}{rlclcrcrl}
\hline\noalign{\smallskip}
&\multicolumn{1}{c}{OGLE} & OGLE & \multicolumn{1}{c}{MACHO}& EROS &\multicolumn{1}{c}{$P_{\rm orb}$}& $V$ & \multicolumn{1}{c}{$V-I$} &\\
\#&\multicolumn{1}{c}{field(s)}& name & \multicolumn{1}{c}{name}& name & \multicolumn{1}{c}{[d]}& [mag] & \multicolumn{1}{c}{[mag]}& Remarks\\
\noalign{\smallskip}\hline\noalign{\smallskip}
   1&LMC\_SC13 & 05065201$-$6825466 & 19.4302.319 &      &  6.33000\hbox{\hspace{1ex}} & 14.036 & $-$0.207 & E, AM \\
  2*&LMC\_SC2  & 05303928$-$7014097 & \hbox{\hspace{1ex}}7.8147.14  &      & 24.673107 & 14.231 & $-$0.041 & E \\
  3*&LMC\_SC11 & 05082813$-$6848251 & \hbox{\hspace{1ex}}1.4539.37  &      &  2.995464 & 14.520 & $-$0.125 & E, AM \\
  4*&LMC\_SC17 & 05381862$-$7041084 & 11.9351.15  &      &  2.191363 & 14.523 &    0.087 & \\
  5*&LMC\_SC7  & 05180481$-$6948189 & 78.6097.13  &      &  3.107029 & 14.591 & $-$0.193 & E \\
  6*&LMC\_SC11 & 05092929$-$6855028 & 79.4779.34  &      &  2.678832 & 14.819 & $-$0.159 & HV\,5622, E, AM \\
  7*&LMC\_SC16 & 05351775$-$6943187 & 81.8881.47  &      &  3.881843 & 14.902 & $-$0.156 & E, AM \\
  8*&LMC\_SC1/16 & 05350218$-$6944178 & 81.8881.44  &      &  2.989472 & 14.941 & $-$0.178 & E \\
  9*&LMC\_SC13 & 05063443$-$6825442 & 19.4302.345 &      &  2.154488 & 15.036 & $-$0.202 & \\
 10*&LMC\_SC7  & 05182818$-$6937453 & 78.6220.60  & 1017 &  1.403789 & 15.094 & $-$0.139 & \\
 11*&LMC\_SC2  & 05322529$-$6925374 & \multicolumn{1}{c}{---}&      &  3.370160 & 15.167 & $-$0.122 & E \\
 12*&LMC\_SC6  & 05210081$-$6929449 & 78.6585.50  &      &  1.300791 & 15.227 & $-$0.196 & \\
 13*&LMC\_SC15 & 05014027$-$6851060 & \hbox{\hspace{1ex}}1.3449.27   &      &  4.034834 & 15.267 & $-$0.105 & E, AM \\
 14*&LMC\_SC5  & 05224434$-$6931435 & 78.6827.52  &      &  2.150539 & 15.391 & $-$0.166 & \\
 15&LMC\_SC18 & 05404706$-$7036564 & 11.9836.21  &      &  7.087663 & 15.420 &    0.075 & \\
 16*&LMC\_SC9  & 05134140$-$6932455 & \multicolumn{1}{c}{---}&      &  5.457320 & 15.437 & $-$0.099 & E\\
 17*&LMC\_SC5/6  & 05223546$-$6931434 & 78.6827.66  & 1036 &  2.183363 & 15.469 & $-$0.178 & \\
 18*&LMC\_SC7  & 05185897$-$6935495 & 78.6221.90  & 1074 &  9.144018 & 15.596 & $-$0.155 & E \\
 19*&LMC\_SC16 & 05371417$-$7020015 & 11.9235.33  &      &  3.256696 & 15.606 & $-$0.069 & HV\,5963, E \\
 20*&LMC\_SC8  & 05164453$-$6932333 & 78.5859.100 & 1066 &  5.603540 & 15.624 & $-$0.143 & E \\
 21*&LMC\_SC10 & 05102875$-$6920480 & \hbox{\hspace{1ex}}5.4894.3904 &      &  3.773443 & 15.632 & $-$0.133 & E, AM \\
 22&LMC\_SC8  & 05170507$-$6945234 & 78.5977.2715&      &  2.864425 & 15.653 & $-$0.180 & E \\
 23*&LMC\_SC11 & 05093433$-$6854259 & 79.4779.81  &      &  1.462906 & 15.663 & $-$0.100 & \\
 24&LMC\_SC13 & 05064333$-$6836115 & 19.4300.349 &      &  4.018698 & 15.672 & $-$0.084 & E, AM \\
 25&LMC\_SC5  & 05225771$-$6935094 & 78.6947.2732&      & 10.187024 & 15.680 & $-$0.146 & \\
 26*&LMC\_SC18 & 05410194$-$7005047 & 76.9844.26  &      &  2.385189 & 15.722 &    0.086 & \\
 27*&LMC\_SC9  & 05132398$-$6922492 & \hbox{\hspace{1ex}}5.5377.4567 &      &  2.636570 & 15.804 &    0.043 & E \\
 28&LMC\_SC10 & 05110289$-$6913098 & 79.5017.83  &      &  2.152919 & 15.923 & $-$0.061 & \\
 29&LMC\_SC6/7  & 05195816$-$6928239 & 78.6465.173 & 1012 &  1.338312 & 15.979 & $-$0.092 & \\
 30&LMC\_SC4  & 05251673$-$6929039 & 77.7312.90  &      &  3.288241 & 16.005 & $-$0.178 & \\
 31&LMC\_SC1  & 05345582$-$6943100 & 81.8881.97  &      &  1.601593 & 16.028 & $-$0.122 & \\
 32&LMC\_SC11 & 05081572$-$6929044 & \hbox{\hspace{1ex}}5.4529.1362 &      &  2.899285 & 16.039 & $-$0.081 & HV\,5619\\
 33&LMC\_SC9  & 05133633$-$6922416 & \hbox{\hspace{1ex}}5.5377.4656 &      &  2.108729 & 16.140 & $-$0.119 & E, AM \\
 34&LMC\_SC5  & 05233948$-$6943346 & 77.7066.333 &      &  2.145747 & 16.194 & $-$0.100 & \\
 35&LMC\_SC15 & 05000218$-$6931561 & 17.3197.781 &      &  3.317560 & 16.245 & $-$0.034 & E \\
 36&LMC\_SC11 & 05095412$-$6853046 & 79.4780.100 &      & 16.930492 & 16.247 & $-$0.051 & E \\
 37&LMC\_SC14 & 05050424$-$6857588 & \hbox{\hspace{1ex}}1.4052.2379 &      &  9.844484 & 16.299 & $-$0.130 & E \\
 38&LMC\_SC8  & 05152057$-$6852502 & 79.5627.88  &      &  1.538050 & 16.301 & $-$0.113 & \\
 39&LMC\_SC4/5  & 05250946$-$7004226 & 77.7303.152 &      &  3.625522 & 16.364 & $-$0.066 & E \\
 40&LMC\_SC18 & 05404159$-$6959014 & 76.9845.63  &      &  2.009981 & 16.367 & $-$0.040 & E \\
 41&LMC\_SC6  & 05221179$-$6928551 & 78.6828.213 &      &  7.544703 & 16.479 & $-$0.166 & E \\
 42&LMC\_SC5  & 05250140$-$6955086 & 77.7184.162 &      &  8.013626 & 16.482 &    0.050 & E \\
 43&LMC\_SC15 & 05010991$-$6904496 & 18.3325.230 &      &  9.771474 & 16.525 & $-$0.045 & E \\
 44&LMC\_SC19 & 05425713$-$7009580 & 76.10205.471&      &  7.229139 & 16.557 &    0.014 & E \\
 45&LMC\_SC6  & 05213496$-$6925346 & 78.6708.180 &      &  2.598897 & 16.565 & $-$0.152 & E, AM \\
 46&LMC\_SC2  & 05315853$-$6955320 & \multicolumn{1}{c}{---}&      &  2.918065 & 16.567 &    0.063 & E \\
 47&LMC\_SC13 & 05053815$-$6820531 & 19.4062.914 &      &  5.621099 & 16.569 & $-$0.122 & E \\
 48&LMC\_SC13 & 05072467$-$6829325 & 19.4423.464 &      &  1.672507 & 16.589 & $-$0.095 & \\
 49&LMC\_SC9  & 05130842$-$6908018 & 79.5260.94  &      &  1.240540 & 16.655 &    0.022 & \\
 50&LMC\_SC10 & 05121954$-$6914547 & 79.5137.189 &      &  8.523814 & 16.697 &    0.086 & E \\
 51&LMC\_SC4  & 05265371$-$6959493 & 77.7546.303 &      &  2.431105 & 16.700 &    0.077 & \\
 52&LMC\_SC4  & 05263667$-$6951253 & 77.7548.325 &      &  2.503626 & 16.702 & $-$0.059 & E, AM \\
 53&LMC\_SC9  & 05134039$-$6918217 & 79.5378.261 &      &  0.956442 & 16.704 & $-$0.088 & \\
 54&LMC\_SC4  & 05263256$-$6945127 & \multicolumn{1}{c}{---}& 1037 &  2.233238 & 16.718 & $-$0.023 & \\
\noalign{\smallskip}\hline
\end{tabular}
\end{table*}

\setcounter{table}{0}
\begin{table*}
\centering
\caption{Continued.}
\begin{tabular}{rlclcrcrl}
\hline\noalign{\smallskip}
&\multicolumn{1}{c}{OGLE} & OGLE & \multicolumn{1}{c}{MACHO}& EROS &\multicolumn{1}{c}{$P_{\rm orb}$}& $V$ & \multicolumn{1}{c}{$V-I$} &\\
\#&\multicolumn{1}{c}{field(s)}& name & \multicolumn{1}{c}{name}& name & \multicolumn{1}{c}{[d]}& [mag] & \multicolumn{1}{c}{[mag]}& Remarks\\
\noalign{\smallskip}\hline\noalign{\smallskip}
 55&LMC\_SC9  & 05130354$-$6917122 & 79.5258.87  &      &  3.289306 & 16.740 & $-$0.055 & \\
 56&LMC\_SC9  & 05143268$-$6912269 & 79.5501.310 &      &  1.232396 & 16.757 & $-$0.091 & \\
 57&LMC\_SC7  & 05174797$-$6904161 & \multicolumn{1}{c}{---}&      &  3.954434 & 16.777 &    0.041 &  \\
 58&LMC\_SC7  & 05174804$-$6945493 & 78.6097.233 &      &  5.231974 & 16.814 & $-$0.088 & E \\
 59&LMC\_SC14 & 05025511$-$6853029 & \hbox{\hspace{1ex}}1.3691.210  &      &  2.789120 & 16.821 & $-$0.089 & E, AM\\
 60&LMC\_SC21 & 05222806$-$7022407 & \hbox{\hspace{1ex}}6.6814.103  &      &  5.376945 & 16.852 &    0.008 & E \\
 61&LMC\_SC4  & 05263908$-$6936060 & 77.7552.249 &      &  3.932107 & 16.852 & $-$0.128 & E \\
 62&LMC\_SC17 & 05383042$-$6949483 & 81.9484.91  &      &  4.261134 & 16.853 & 0.031 & E, AM \\
 63&LMC\_SC1  & 05324193$-$6951092 & 81.8516.176 &      &  4.954343 & 16.892 & $-$0.064 & E, AM \\
 64&LMC\_SC4  & 05260950$-$6959109 & 77.7425.249 &      & 17.366232 & 16.917 & $-$0.027 & E \\
 65&LMC\_SC9  & 05132215$-$6927252 & \hbox{\hspace{1ex}}5.5376.2159 &      &  1.788358 & 16.927 & $-$0.106 & \\
 66&LMC\_SC6  & 05222482$-$6936226 & 78.6826.298 & 1041 &  2.688218 & 16.929 &    0.003 & E \\
 67&LMC\_SC2  & 05312473$-$6925281 & 77.8281.58  &      &  2.536663 & 16.944 & $-$0.036 & E \\
 68&LMC\_SC11 & 05092769$-$6856194 & 79.4658.4032&      &  1.827980 & 16.983 & $-$0.079 & \\
 69&LMC\_SC3  & 05285710$-$6948441 & 77.7912.282 &      &  6.069953 & 16.992 & $-$0.044 & \\
 70&LMC\_SC9  & 05142797$-$6854210 & 79.5505.198 &      &  9.002650 & 16.995 &    0.078 & E \\
 71&LMC\_SC5  & 05224192$-$7006480 & \hbox{\hspace{1ex}}6.6818.283  &      &  4.210379 & 17.001 & $-$0.023 & \\
 72&LMC\_SC6  & 05201732$-$7000440 & \hbox{\hspace{1ex}}6.6457.4965 &      &  2.116275 & 17.021 & $-$0.036 & \\
 73&LMC\_SC14 & 05041689$-$6849509 & \hbox{\hspace{1ex}}1.3812.152  &      &  5.280710 & 17.029 &    0.016 & E \\
 74&LMC\_SC9  & 05133011$-$6908412 & 79.5381.199 &      &  9.503056 & 17.036 &    0.018 & E\\
 75&LMC\_SC3  & 05292500$-$6948081 & 77.7912.709 &      &  1.966261 & 17.067 & $-$0.058 & \\
 76&LMC\_SC13 & 05062656$-$6857153 & \hbox{\hspace{1ex}}1.4174.183  &      &  5.322229 & 17.094 &    0.013 & E \\
 77&LMC\_SC6  & 05211299$-$6950512 & 78.6580.255 &      &  2.160515 & 17.107 & $-$0.101 & E \\
 78&LMC\_SC10 & 05121869$-$6858325 & 79.5141.200 &      &  2.390521 & 17.113 &    0.005 & E \\
 79&LMC\_SC10 & 05122789$-$6920513 & \hbox{\hspace{1ex}}5.5257.3679 &      &  1.982183 & 17.151 &    0.106 & \\
 80&LMC\_SC9  & 05140857$-$6923003 & \hbox{\hspace{1ex}}5.5498.5030 &      &  1.628412 & 17.164 & $-$0.107 & \\
 81&LMC\_SC7  & 05181228$-$6936251 & 78.6100.388 & 1061 &  4.538132 & 17.182 & $-$0.131 & E \\
 82&LMC\_SC1  & 05331282$-$7007025 & 81.8512.224 &      &  5.394376 & 17.211 & $-$0.037 & E \\
 83&LMC\_SC5/6  & 05223386$-$6932564 & 78.6827.465 &      &  1.284099 & 17.224 & $-$0.145 & \\
 84&LMC\_SC6  & 05221500$-$6938483 & 78.6825.430 & 1063 &  4.722907 & 17.226 & $-$0.055 & E \\
 85&LMC\_SC6  & 05203518$-$6934378 & 78.6463.505 &      &  2.117483 & 17.226 & $-$0.142 & E \\
 86&LMC\_SC6  & 05222499$-$6938103 & 78.6825.431 & 1053 &  3.570088 & 17.259 & $-$0.180 & E \\
 87&LMC\_SC8  & 05170646$-$6940570 & 78.5978.403 &      &  1.635801 & 17.268 & $-$0.042 & \\
 88&LMC\_SC10 & 05112066$-$6909148 & 79.5018.187 &      &  1.345239 & 17.278 &    0.035 & \\
 89&LMC\_SC10 & 05115154$-$6920494 & 79.5136.250 &      &  1.760103 & 17.280 & $-$0.005 & E, AM \\
 90&LMC\_SC4  & 05264527$-$6944045 & 77.7550.352 &      &  6.536197 & 17.304 & $-$0.024 & E \\
 91&LMC\_SC7  & 05181122$-$6932555 & 78.6101.407 &      &  3.816795 & 17.317 & $-$0.069 & \\
 92&LMC\_SC6  & 05222991$-$6919090 & 80.6830.375 &      &  6.310800 & 17.363 & $-$0.028 & E \\
 93&LMC\_SC1  & 05335582$-$7019049 & 11.8630.374 &      &  4.568919 & 17.368 &    0.112 & \\
 94&LMC\_SC9  & 05145023$-$6915416 & 79.5621.470 &      &  4.670019 & 17.399 & $-$0.018 & \\
 95&LMC\_SC2  & 05323120$-$6928535 & 81.8522.169 &      &  4.897253 & 17.431 &    0.054 & E \\
 96&LMC\_SC7  & 05181271$-$6935245 & 78.6100.606 & 1039 &  2.575579 & 17.460 & $-$0.032 & E \\
 97&LMC\_SC6  & 05221305$-$7003284 & 78.6819.336 &      &  2.138851 & 17.466 &    0.019 & E \\
 98&LMC\_SC14 & 05025406$-$6918398 & \hbox{\hspace{1ex}}1.3684.237  &      &  3.825716 & 17.475 & $-$0.037 & \\
\noalign{\smallskip}\hline
\end{tabular}
\end{table*}

\section{New transformation to magnitudes}

As mentioned in Sect.~2.1, two sources of time-series data in the $I$ filter expressed
in magnitudes are available
from the OGLE-II survey.  At the beginning, we had to decide which of them use in the
subsequent analysis.  Due to the crowded nature of the LMC fields, the DIA photometry
has considerably smaller scatter than that obtained
by means of profile fitting and hence it seems to be preferable.  However,
this photometry suffers from the unavoidable bias occuring during the
transformation from differential fluxes to magnitudes (see, e.g., Wo\'zniak \cite{wozniak} or
\.Zebru\'n et al.~\cite{zebrun}).  The bias in the transformed magnitude, $m_{\rm DIA}$, comes mainly
 from the uncertainty of
the reference flux, $f_{\rm ref}$, in the transformation equation:
\begin{equation}
m_{\rm DIA}=-\mbox{2.5}\log(f_{\rm ref} + \Delta f) + C.
\end{equation}
In the above equation, $\Delta f$ denotes differential flux, and $C$ is the zero point of the
magnitude scale.  While the uncertainty in $C$ does not affect the shape and the magnitude range of
the transformed light curve, that of $f_{\rm ref}$ does.  In consequence, the parameters derived
from the fit of the light curve would be affected as well.

The problem is illustrated in Fig.~\ref{dia-dophot} for a star that was observed within OGLE-II in two
overlapping fields.  The top panel shows the light curve derived by means of {\sc DoPhot}.
The light curves from the two fields agree quite well, there is only a small 0.003-mag shift in the mean
brightness between them.  On the other hand, the DIA light curves (middle panel) are less scattered, but
one is shifted by $\sim$0.1~mag with respect to the other and the eclipses have slightly different depths.  This is
because $f_{\rm ref}$ was derived merely from the profile-fitting in the reference images.  The reference images,
although obtained by co-adding 20 best frames in each field (\.Zebru\'n et al.~\cite{zebrun}),
were still quite crowded.
Therefore, we calculated new $f_{\rm ref}$ and $C$ values in Eq.~(1), fitting the DIA differential fluxes to the {\sc DoPhot}
magnitudes by means of least-squares fits.  The result is shown in Fig.~\ref{dia-dophot}c.
This approach combines the advantages of both photometries: small scatter of the
DIA photometry and the correct range in brightness of the {\sc DoPhot} one.  A new transformation
was derived for all eclipsing binaries selected for further analysis.  Although the example shown in
Fig.~\ref{dia-dophot} is an extreme rather than a typical case, we encourage all users of the 
DIA photometry to do a new transformation in all applications where the correct magnitude scale is 
required.

\section{Analysis of light curves}
The light curves in three (or four, if the EROS $B_{\rm E}$ photometry was available) bands,
MACHO blue and red, (designated thereafter $V_{\rm M}$
and $R_{\rm M}$, respectively) and OGLE $I$, were analyzed simultaneously by means of
the improved version of the Wilson-Devinney (WD) program that includes a model atmosphere routine
developed by Milone et al.~(\cite{milone}).  The program is composed of the LC program generating
light curves for assumed parameters and the differential correction program, DC.  The WD program
was used in two different ways: (1) to search the space of the adjustable parameters by means of
the Monte-Carlo method with the LC program, (2) to get the best-fit solution for a given system
with the DC one.

\subsection{Assumptions}

The DC program allows adjusting of over thirty parameters.  In practice, only a few are adjusted
since the remaining are either known quite well or cannot be reliably obtained from the fit.
The choice of adjustable parameters have to be decided first.

Since the components of the DEBs are roughly spherical in shape, it is not possible to get the
mass ratio, $q = M_2/M_1$, solely from the analysis of the light
curves (Wyithe \& Wilson \cite{WW1}).  Michalska \& Pigulski (\cite{our-prel})
presented an example showing that equally good fits can be obtained in a very large
range of mass ratios.  It is therefore reasonable to assume $q$ = 1, as we did in this analysis,
although it is obvious that for systems with unequal minima, $q$ might be far from
unity.  The WD program was run with detached geometry (MODE = 2), as the selected stars
were suspected to be DEBs.  If, however, the gravitational potential of one of the
components appeared to be close to the critical one, the system was excluded from the
analysis.

The choice we made in Sect.~3 allowed us to assume safely that the components of the
analyzed systems were early-type main-sequence stars.  Consequently, we assumed bolometric albedos
and gravity darkening coefficients equal to 1.0, a value typical for stars with radiative envelopes.
The logaritmic limb darkening law (LD = 2) with coefficients taken from van Hamme
(\cite{hamme}) for $\log g$ = 4.0 was adopted.  In addition, we assumed synchronous rotation and no spots.
As far as the reflection effect is concerned, a detailed reflection
model with two reflections (MREF = 2 and NREF = 2) was assumed for systems with circular orbits.
For eccentric systems a simpler model with single reflection (MREF = 1 and NREF = 1) was adopted.
This difference is not important as the reflection effect is never significant for the systems we selected.

The major semi-axis of the relative orbit, $a$, has little to do with the light curve.
However, since even an estimation gives an idea of the expected range of radial velocities,
we tried to evaluate $a$.  At the beginning, it was assumed to be equal to 25~$R_\odot$. Once the masses
of the components were estimated (see below), $a$ was calculated from the generalized Kepler's law.

\begin{figure}
\includegraphics[width=8.7cm,clip]{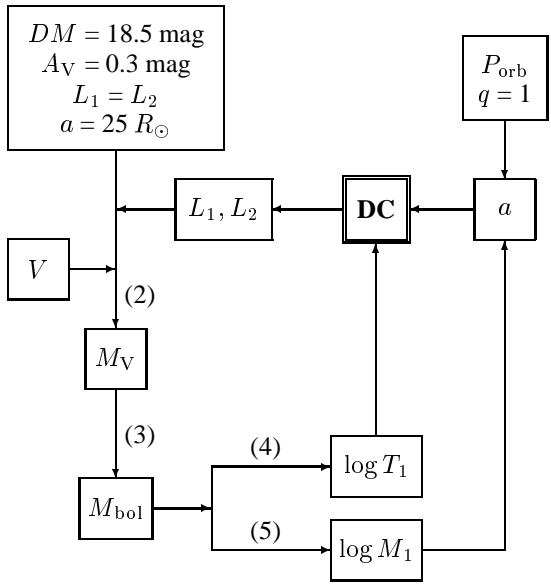}
\caption{The scheme of the procedure aimed in estimating the effective temperature of the primary, $T_1$.
See Sect.~5.1 for details.}
\label{schem}
\end{figure}

The effective temperature of the primary component, $T_1$, is the last important parameter
that needs to be assumed.
In order to do this in a consistent way, we employed an iterative procedure shown schematically
in Fig.~\ref{schem}.  Using the out-of-eclipse $V$ magnitudes, we first estimated the absolute magnitude
of the primary, $M_{\rm V}$, according to the equation:
\begin{equation}
M_{\rm V} = V - DM - A_{\rm V} - \mbox{2.5} \log \frac{L_1}{L_1+L_2},
\end{equation}
where the average values for the LMC distance modulus, $DM$ = 18.5~mag, and total extinction, $A_{\rm V}$ = 0.3~mag,
were assumed.  The component's monochromatic luminosities were first assumed to be the same, $L_1$ = $L_2$,
then they were taken from the DC output (see Fig.~\ref{schem}).
Having calculated $M_{\rm V}$, the bolometric magnitude, $M_{\rm bol}(M_{\rm V})$ relation
in a form of a sixth-order polynomial
was derived using the bolometric corrections published by Popper (\cite{popper}).
The coefficients, $a_n$, of the polynomial
$$P(A) = \sum\limits_{n=0}^{N} a_n A^n,$$
are given in Table \ref{poly}.

\begin{table}
\caption{The coefficients of the polynomials in the relations (3)--(5). $P$ is a parameter, $A$, the argument of a relation.}
\label{poly}
\begin{tabular}{rrrr}
\hline\noalign{\smallskip}
Eq. & (3) & (4) & (5) \\
$P(A)$ & $M_{\rm bol}(M_{\rm V})$ & $\log M_1(M_{\rm bol})$ & $\log T_1(M_{\rm bol})$ \\
\noalign{\smallskip}\hline\noalign{\smallskip}
$N$ & 6 & 4 & 6 \\
$a_0$& $-$1.46072 &      0.468944 &       4.07891 \\
$a_1$&    1.67977 &  $-$9.46798 $\cdot$ 10$^{-2}$ &  $-$8.02904 $\cdot$ 10$^{-2}$\\
$a_2$& 1.94271 $\cdot$ 10$^{-2}$ & $-$4.76657 $\cdot$ 10$^{-4}$ &  $-$1.69451 $\cdot$ 10$^{-3}$ \\
$a_3$& $-$3.00150 $\cdot$ 10$^{-2}$ &  $-$1.16486 $\cdot$ 10$^{-4}$ & 1.16818 $\cdot$ 10$^{-3}$ \\
$a_4$& $-$1.53159 $\cdot$ 10$^{-3}$ &  4.00185 $\cdot$ 10$^{-5}$ &  9.66154 $\cdot$ 10$^{-5}$ \\
$a_5$& 1.07280 $\cdot$ 10$^{-3}$ &  ... & $-$3.35212 $\cdot$ 10$^{-5}$ \\
$a_6$& $-$7.38667 $\cdot$ 10$^{-5}$ & ... &  $-$3.48540 $\cdot$ 10$^{-6}$ \\
\noalign{\smallskip}\hline
\end{tabular}
\end{table}

The coefficients of the effective temperature vs.~luminosity, $\log T_1(M_{\rm bol})$, and
mass-luminosity, $\log M_1(M_{\rm bol})$ relations, are also given in Table \ref{poly}.
They were derived from the data published for Galactic detached main-sequence
eclipsing binaries by Harmanec (\cite{harmanec}).  In order to account for smaller
metallicity of the LMC, we applied a correction taken from the comparison of stellar
parameters for $Z$ = 0.02 and $Z$ = 0.008 models of Bertelli et al.~(\cite{bertelli})
for the youngest stars ($\log({\rm age/year})$ = 6.6,  $M_1 <$ 15\,$M_\odot$).

\subsection{Adjusted parameters}

The adjusted parameters were the following:
\begin{itemize}
\item[$\bullet$] phase shift, $\phi_0$,
\item[$\bullet$] surface potentials, $\Omega_1$ and $\Omega_2$,
\item[$\bullet$] effective temperature of the secondary component, $T_2$,
\item[$\bullet$] inclination, $i$,
\item[$\bullet$] luminosity of the primary component, $L_1$.
\end{itemize}
For a few systems (\#21, \#28, \#29, \#32, \#66, \#73, \#81, and \#88) it was necessary to include into the solution the third light,
$L_3$, as an additional adjustable parameter.
The DC program provides also the fractional radii, $r_1$ and $r_2$ (back, pole, point, and side),
and the luminosity of the secondary, $L_2$.  After a first guess of the parameters,
the procedure shown in Fig.~\ref{schem} was repeated five times.  Then, $a$ and
$T_1$ were fixed and the iterations were repeated until the solution converged.
In order to scrutinize the solution, we utilized the Monte-Carlo method (Sect.~5.3) that
warrants finding a global minimum in the parameter space.
\begin{figure*}
\includegraphics[width=508pt,clip]{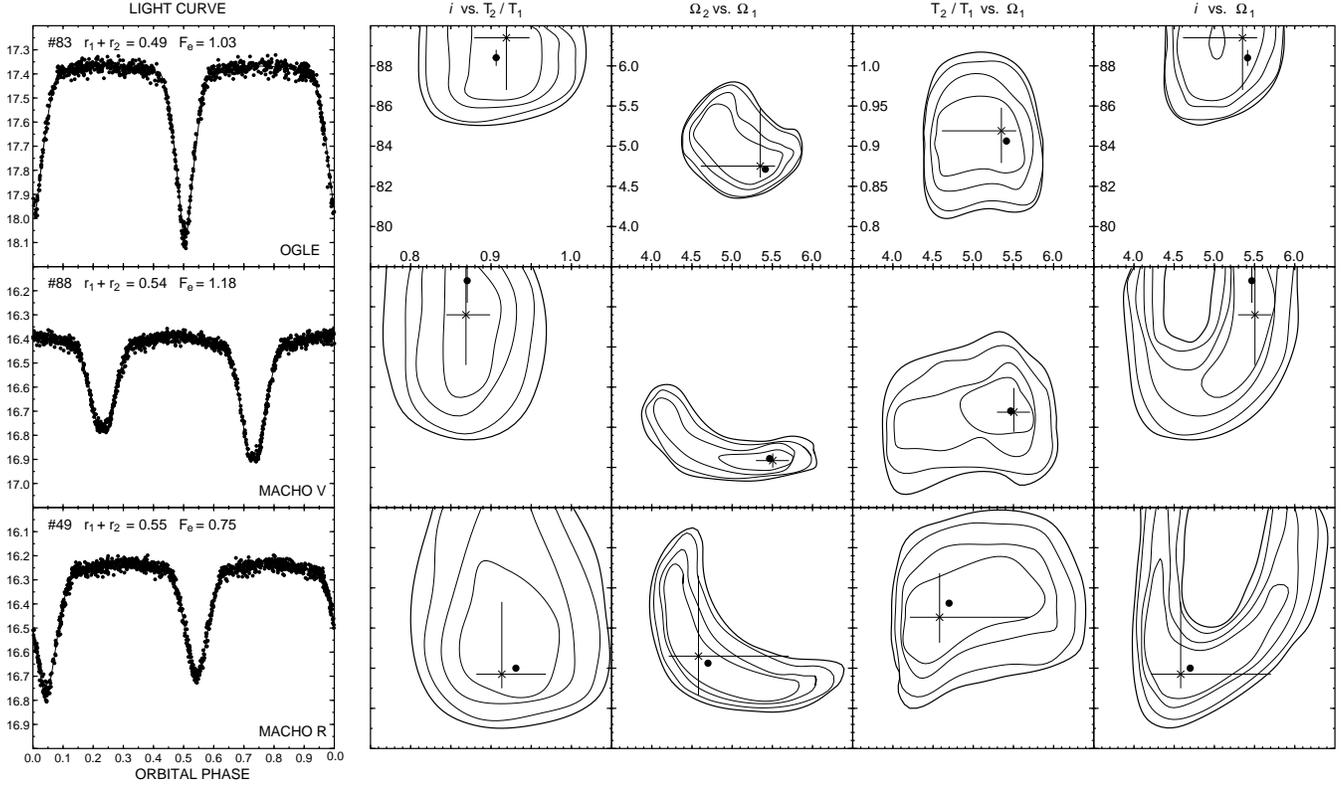}
\caption{Results of the M-C simulations for three DEBs with different light-curve morphology. {\it Left:} Light
curves of \#83 (complete eclipses, similar components), \#88 (complete eclipses, different components, and \#49
(partial eclipses).  The ordinate range is the same for all three stars and equals to 1~mag.
{\it Right:} Panels showing the dependencies between $i$, $T_2/T_1$, $\Omega_1$ and $\Omega_2$.  The
ranges of the parameters are the same as labeled at the top.  The contours we show encompass all solutions
with SSR smaller than $A$ $\cdot$ SSR$_{\rm min}$, for four values of $A$: 2, 1.73, 1.41, and 1.22.
The WD and M-C solutions are plotted as dots and crosses, respectively, with error bars.}
\label{mcs}
\end{figure*}

\subsection{Monte-Carlo simulations}
The Monte-Carlo (M-C) method is often used in problems with a complicated parameter space and
strong correlations between parameters.  It also provides an independent way to estimate
the uncertainties of the parameters in the presence of the mentioned correlations.

The preliminary solution, obtained by means of the WD program, was used to define the
ranges of parameters that were searched for the best solution.  The parameters were the same
as listed in Sect.~5.2, except for $\phi_0$ that was fixed at the value
obtained during the DC iterations.  The synthetic light curves were
calculated using the LC program with a randomly-generated (within a defined range)
set of parameters.
The weighted sum of squares of the residuals (hereafter SSR)
was stored as the goodness-of-fit estimator.  At this stage we faced the problem of whether to
compare the synthetic and observed light curves in all available bands and combine
the resulting weighted SSRs to get a general estimation of the multi-band solution or do the same for a single band.
The first approach has the advantage of being able to discriminate better solutions in case of unequal
components.  However, when the components are similar and there are large differences in the scatter
between different bands, a single band determined the
shape of the minimum in the parameter space.  For that reason we decided to follow the second approach
(single band), choosing the band with the smallest scatter in photometry.  The band we chose is indicated in the last column
of Table 3.  Only in a few cases, when
the scatter was similar in two bands, we did trial M-C simulations for these two bands separately
and then compared the results.
The global solutions obtained with the M-C method appeared to be
consistent with the preliminary ones for all but a few systems.

Some examples of the results of the M-C simulations are shown in Fig.~\ref{mcs} for three systems
with different light-curve morphology.  In this figure, we plot contours that encompass all solutions
with SSR smaller than a certain value expressed in terms of the minimum value of SSR.
Let us first comment on the $\Omega_2$ vs.~$\Omega_1$ panels, as the surface potentials decide how well
the components' dimensions are constrained.  We see that $\Omega$s are best constrained for
a system with complete eclipses and components that are not similar (\#88).  In the case of similar components
(\#83) we can also get $\Omega$s, but there are two alternative solutions.  This is obvious, as
{\it the same} components in a circular orbit produce eclipses indistinguishable in shape.  Wyithe \& Wilson (\cite{WW1})
call these solutions `aliasing'.  For partial eclipses (\#49), the solutions spread over a crescent-shaped area that
is quite wide in both $\Omega$s.  This is understandable
as the change in $\Omega$ can be easily compensated by
change of inclination.  The relative radii for a system with partial eclipses is therefore much poorly
constrained.  In general, this means that such systems are much less
suitable for the distance determination.
\begin{figure*}
\includegraphics[width=508pt,clip]{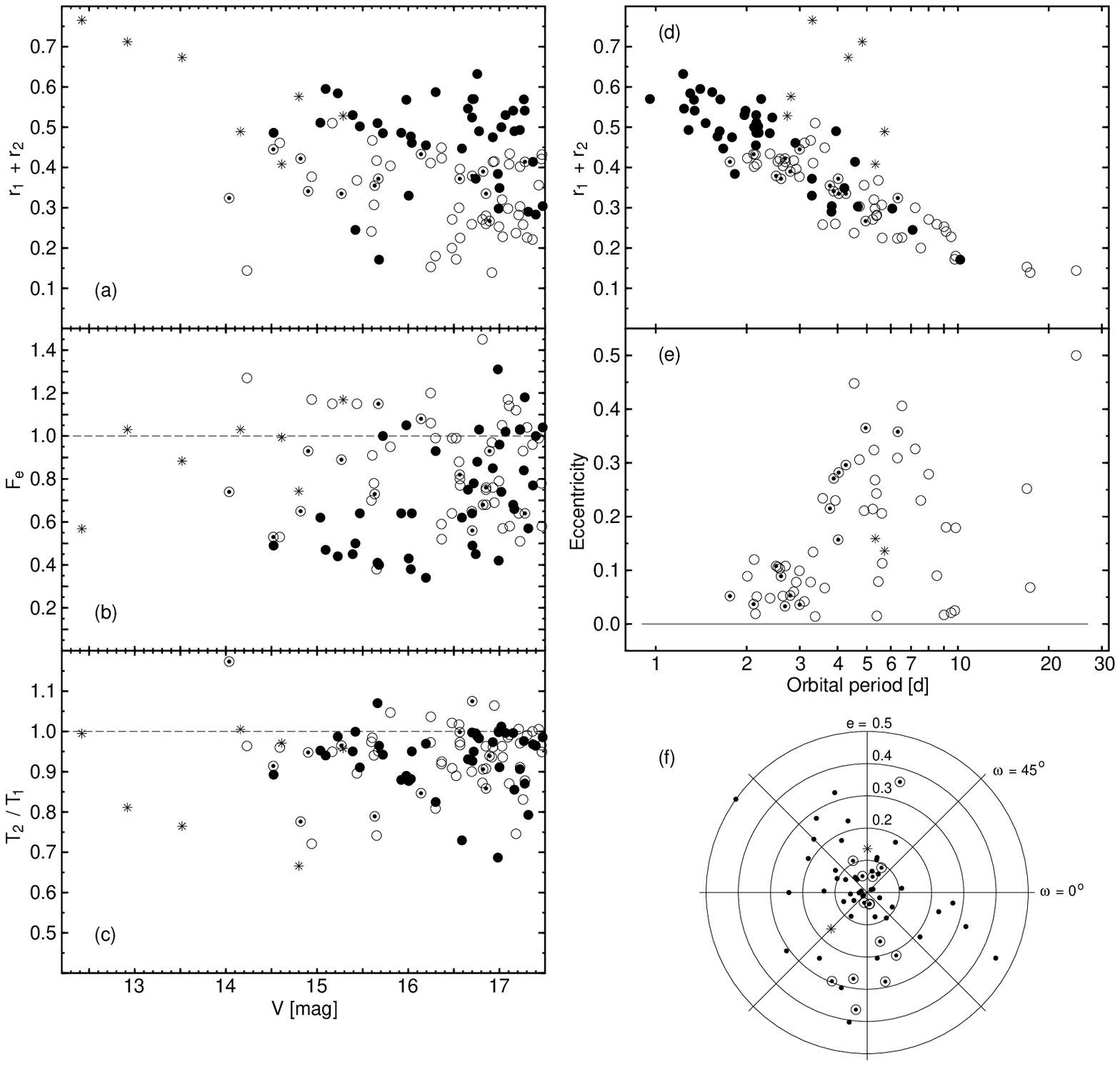}
\caption{Parameters of 98 systems selected in this paper. For comparison, the same parameters are also
plotted (as asterisks) for the seven systems listed in the Introduction that were already used to derive the distance of the LMC.
{\it Left:} Sum of relative radii, $r_1+r_2$ (a), $F_{\rm e}$ parameter
(b), and ratio of effective temperatures, $T_2/T_1$, plotted against the $V$ magnitude.  Parameters for systems
with circular orbits are plotted as filled circles, for eccentric systems, as open circles, and the eccentric systems with
detected apsidal motion, as encircled dots.
{\it Right:} Sum of the relative radii, $r_1+r_2$ (d), and eccentricity, $e$ (e), plotted as a function of the
orbital period.  The same symbols as in panels (a)--(c) are used.  The (f) panel shows the $e$ vs.~$\omega$ (longitude
of periastron) dependence.  Systems with apsidal motion are plotted as encircled dots.}
\label{rfe}
\end{figure*}

\subsection{Discussion of solutions}
Table 3 lists the parameters of 98 systems derived by means of
the WD program.  For 84 stars from this sample we also provide the results obtained
with the M-C simulation.  The remaining fourteen eccentric systems show a detectable apsidal motion and
the M-C simulation was not performed for these stars.
They are discussed separately in Sect.~6.

Since, as shown above, surface potentials are not well constrained for the DEBs, we instead provide in Table 3
the sum of relative radii,
$r_1 + r_2$.  In addition, the $F_{\rm e}$ parameter is given, defined by Wyithe \& Wilson (\cite{WW1}) as:
\begin{equation}
F_{\rm e} = \frac{r_{\rm 1}+r_{\rm 2}-\cos i}{2r_{\rm 2}},
\end{equation}
where $r_{\rm 1}$ and $r_{\rm 2}$ denote the fractional radii (in terms of the relative distance during the eclipse)
of the larger and smaller star, respectively.   For systems with total
eclipses, $F_{\rm e} \geq$ 1.  However, the denominator of (3) is not well known, so that the
values of $F_{\rm e}$ we provide should be treated with caution.  In the case of an eccentric orbit, 
the relative distance can be different in both eclipses.  For this reason we list two values of
$F_{\rm e}$ in Table 3.
Finally, the range of radial velocities, resulting from the derived
parameters, masses of the components and dimensions estimated in Sect.~5.1, are given in the tenth column of Table 3.
This is a rough estimation, but it can be useful in view of the future spectroscopic observations.

The parameters of the systems are also shown in Fig.~\ref{rfe}a-f.  As a result of the application
of the selection criteria (Sect.~3), we deal mostly with systems that have similar components.
This can be seen in Fig.~\ref{rfe}c: for most systems 0.8 $< T_2/T_1 <$ 1.0, with only a few stars outside this
range.  From Fig.~\ref{rfe}a and \ref{rfe}d we see that the closest systems (i.e., having largest $r_1+r_2$)
have---as expected---mostly circular orbits and shortest periods.  For $r_1+r_2 \geq$ 0.5 all orbits are circularized,
which is consistent with the results of North \& Zahn (\cite{noza}). Practically all systems
with $P_{\rm orb}$ shorter than $\sim$2~d have circularized orbits.  This is a value typical for early-type, i.e., young systems,
but much shorter than in older populations (Mathieu \& Mazeh \cite{mama}).
\begin{figure*}
\includegraphics[width=508pt,clip]{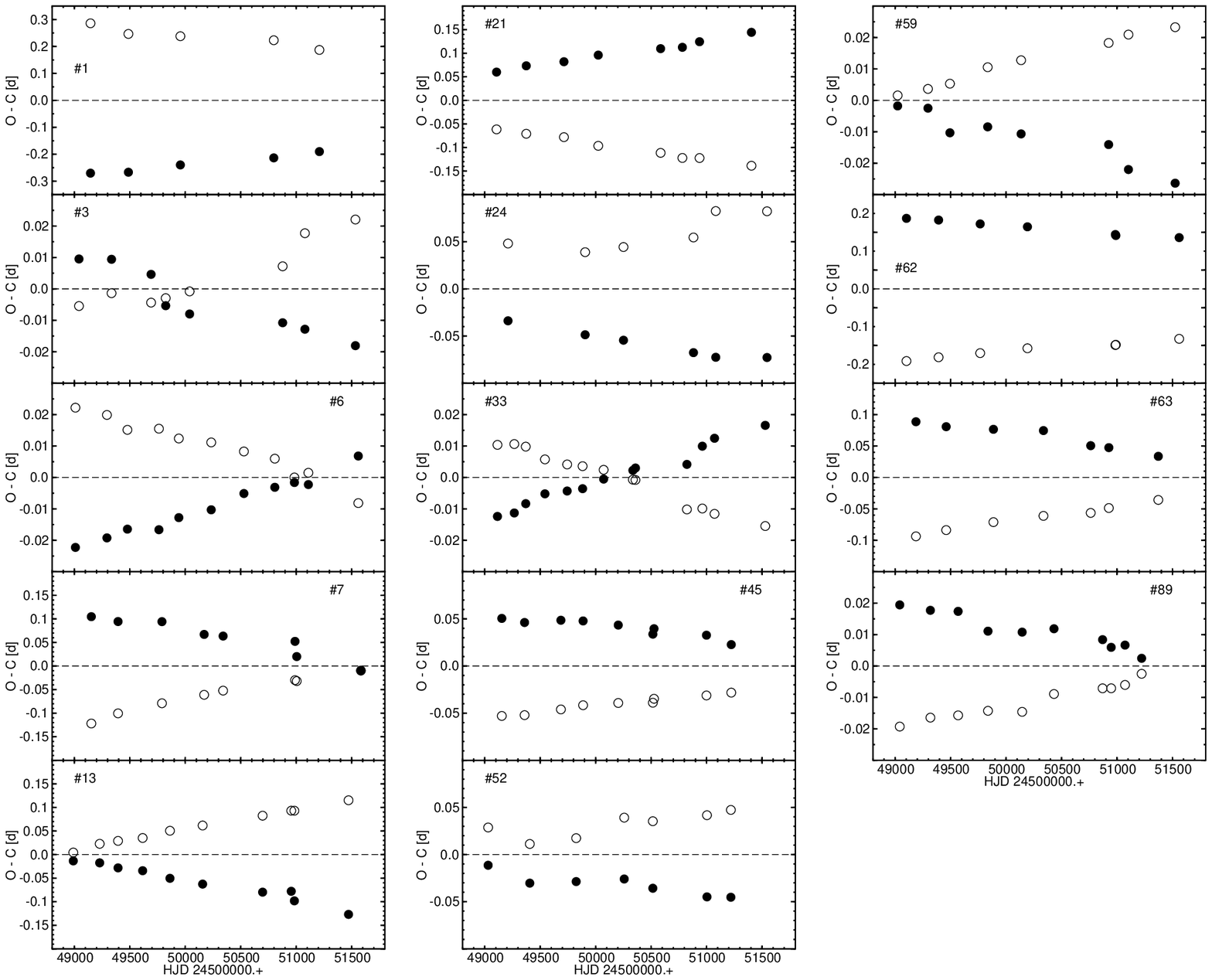}
\caption{The $O$\,$-$\,$C$ diagrams for fourteen systems with apsidal motion.  The filled and open circles denote
the primary and secondary times of minimum, respectively.  Note different ordinate scales.}
\label{apsidal}
\end{figure*}

\section{Systems with apsidal motion}
As a consequence of the tidal and rotational distortions, the apsidal motion is observed in some
eccentric systems.  With the aid of theory, the knowledge of the rate of apsidal motion
can be used to test the internal structure of the binary components (Claret \& Gim\'enez \cite{clgi93},
Claret \cite{claret} and references therein)
or even derive their masses (Benvenuto et al.~\cite{benv02}).
We detected that fourteen eccentric systems listed in Table \ref{tab-list} exhibit measurable
apsidal motion.  Since the WD program allows fitting of the rate of change of the longitude of
periastron, $\dot\omega$ = $d\omega/dt$, we derived these values from the WD fits.
They are listed in the fourth column of Table \ref{tab-aps}.  The values of $e$ and $\omega$ are repeated
from Table 3 for reference.  Note that for systems with apsidal motion, $\omega$ is given for
epoch HJD\,2450500.0.
\setcounter{table}{3}
\begin{table}
\centering
\caption{Parameters for 14 systems with apsidal motion detected.  The values of $e$ and $\omega$ were
repeated from Table 3.  The remaining parameters are explained in the text.}
\label{tab-aps}
\begin{tabular}{cccclc}
\hline\noalign{\smallskip}
Star & $e$ & $\omega$&\multicolumn{1}{c}{$\dot\omega$} & \multicolumn{1}{c}{$P_{\rm mean}$} & $T_{\rm 0,\,mean}$ \\
     & &[$^{\rm o}$] &\multicolumn{1}{c}{[$^{\rm o}$/year]} & \multicolumn{1}{c}{[d]} & [HJD\,244...] \\
\noalign{\smallskip}\hline\noalign{\smallskip}
\#1 & 0.358 & \hbox{\hspace{1ex}}74 & 0.86 $\pm$ 0.04 & 6.33000 & 9146.1025 \\
\#3 & 0.036 & 283 & 6.9 $\pm$ 1.0 & 2.995459 & 9041.9927 \\
\#6 & 0.033 & 254 &7.2 $\pm$ 0.8 & 2.6788316 & 9009.9189\\
\#7 & 0.271 & 261 &2.35 $\pm$ 0.06 & 3.881843 & 9153.4719\\
\#13 & 0.282 & 282 &2.35 $\pm$ 0.04 & 4.034834 & 8890.6211\\
\#21 & 0.215 & 295 &3.05 $\pm$ 0.07 & 3.773443 & 9106.6226\\
\#24 & 0.157 & 285 &2.05 $\pm$ 0.10 & 4.018700 & 9209.9996\\
\#33 & 0.037 & 280 & 8.7 $\pm$ 0.7 & 2.1087292 & 9114.7029\\
\#45 & 0.089 & \hbox{\hspace{1ex}}60 & 3.64 $\pm$ 0.28 & 2.5988972 & 9153.9262\\
\#52 & 0.108 & 114 & 3.31 $\pm$ 0.23 & 2.503626 & 9030.7418\\
\#59 & 0.053 & 106 & 3.2 $\pm$ 0.4 & 2.7891195 & 9022.6667\\
\#62 & 0.296 & 248 & 1.21 $\pm$ 0.06 & 4.261134 & 9102.1955\\
\#63 & 0.365 & 265 & 0.79 $\pm$ 0.03 & 4.954343 & 9187.7252\\
\#89 & 0.052 & \hbox{\hspace{1ex}}71 & 5.9 $\pm$ 0.6 & 1.7601028 & 9041.5233\\
\noalign{\smallskip}\hline
\end{tabular}
\end{table}

The $O$\,$-$\,$C$ diagrams for the 14 systems under consideration are shown in Fig.~\ref{apsidal}.
They were obtained in the following way.  First, the MACHO blue and OGLE-II data were
divided into five to thirteen subsets, depending on the star.  Then, the WD program was
run separately for each subset with $e$ and $\omega$ fixed.  The phase of primary and secondary
minimum was next derived from the fit.  These phases were transformed into
times of minimum closest to the mean epoch of all observations in a given subsample.

Having obtained the times of minimum from all subsets, we calculated the
number of elapsed cycles, $E$, for each minimum and then fitted them with a function
$$T_{\rm min} = T_0 + P \times E$$
deriving $T_0$ and $P$, separately for the primary and secondary ones.
A mean of both periods, $P_{\rm mean}$ = ($P_{\rm prim}$ + $P_{\rm sec})/$2,
is given in Table \ref{tab-aps}. We have also averaged the initial epochs;
the mean initial epoch $T_{\rm 0,\,mean}$ given in Table \ref{tab-aps} was equal to the average of $T_0^{\rm prim}$ and
$T_0^{\rm sec}$ + $P_{\rm mean}/$2.  Using the observed epochs, the values of $P_{\rm mean}$ and $T_{\rm 0,\,mean}$,
we calculated the $O$\,$-$\,$C$ values plotted in Fig.~\ref{apsidal}, where the calculated epochs $C$ were
derived from the equation
\begin{equation}
C = T_{\rm 0,\,mean} + P_{\rm mean} \times E
\end{equation}
for the primary and
\begin{equation}
C = T_{\rm 0,\,mean} + P_{\rm mean} \times (E + 0.5)
\end{equation}
for the secondary minima.

As can be seen in Fig.~\ref{rfe}e, systems with detected apsidal motion tend to group around $\omega \sim$90$^{\rm o}$ and
270$^{\rm o}$.  This is clearly an observational selection effect as in these configurations the effect on the position of the
observed minima is largest.  The $\dot\omega$ values are not determined very accurately (Table \ref{tab-aps}),
but they can be used to estimate the apsidal periods, $U$ = 360$^{\rm o}/\dot\omega$.
The values of $U$ are generally longer than $\sim$40 years.  The lack of systems with shorter $U$ can be understood
as we selected only the DEBs. In DEBs the tidal effects are not as strong as in the systems with closer components
for which $U$ is sometimes very short (Petrova \& Orlov \cite{peor}).

The systems with apsidal motion we found are not the first known in the LMC.  Apsidal motion
was detected in HV\,2274 by Watson et al.~(\cite{watson}) and thereafter studied by Claret (\cite{clar96}).
It was also found in another well-studied DEB in the LMC, HV\,982 (Clausen et al.~\cite{clausen}).
In addition, Bayne et al.~(\cite{bay04}) recently detected this effect in the LMC system MACHO 05:36:48.7$-$69:17:00
= 82.9130.25.

\section{Conclusions}
We have selected 98 detached systems in the LMC that have good photometry available.  They were
extracted from the OGLE-II, MACHO and EROS databases and combined.
For the OGLE-II data we performed transformations from fluxes to magnitudes, ensuring both small scatter and
the correct range of changes.
The next step towards the determination of the distance to the LMC is to obtain radial velocities for at least
some of these stars.  Since they all are relatively bright, this could be achieved even with 4-m class telescopes.

For all systems we derived astrophysical parameters that could be obtained from the fitting of the light-curve.
They can be used to select the targets for spectroscopic observations.
Comparison of the parameters of systems that were already used to derive the distance to the LMC with those we select in this
paper (Fig.~\ref{rfe}) indicates clearly that among the latter there are many systems that are suitable for this purpose.

Special attention should be paid to the fourteen systems in which we detected apsidal motion.  Fortunately, OGLE-III
continues to observe the LMC, so that we can expect that they may be studied in the future in a much more detail.

\begin{acknowledgement}
This work was supported by the KBN grant No.\,1\,P03D\,016\,27.  We thank Z.\,Ko{\l}aczkowski for
providing scripts used in the selection of eclipsing variables.
\end{acknowledgement}

\vfill\eject
\setcounter{table}{2}
\begin{table*}
\centering
\caption{Parameters of the 98 selected systems.  If two lines are presented for a given star, the first one refers to the results of the WD fitting, the other, to the
M-C simulations.  The errors given for the M-C simulations correspond to the 10\% increase of the SSR with respect to the minimum value.  
The columns are the following: (1), sequential number, (2) effective temperature of star 1 [K] (assumed, see Sect.~5.1), (3) effective 
temperature of  star 2 [K], (4) inclination [$^{\rm o}$], (5) sum of relative radii, $r_1 + r_2$, (6),(7) $F_{\rm e}$ parameter; in case of 
eccentric orbits, two values are given: first corresponds to the primary, the other, to the secondary eclipse, (8) eccentricity, (9) longitude of
periastron [$^{\rm o}$], (10) estimated semi-amplitude of the radial-velocity changes [km s$^{-1}$], (11) filter indicating the data used
in the M-C simulations, `O', OGLE-II, `MR', MACHO red, `MB', MACHO blue.}
\label{tab-par}
\begin{tabular}{rclllccclcc}
\hline\noalign{\smallskip}
\#& $T_1^{*}$ & \multicolumn{1}{c}{$T_2$}& \multicolumn{1}{c}{$i$} & \multicolumn{1}{c}{$r_1 + r_2$} & 
   \multicolumn{1}{c}{$F_{\rm e,1}$} & \multicolumn{1}{c}{$F_{\rm e,2}$} & \multicolumn{1}{c}{$e$} & \multicolumn{1}{c}{$\omega$} &
   \multicolumn{1}{c}{$K_{1,2}$} & M-C \\
  &   [K]     & \multicolumn{1}{c}{[K]}  & \multicolumn{1}{c}{[$^{\rm o}$]} &&&&&\multicolumn{1}{c}{[$^{\rm o}$]} & 
   \multicolumn{1}{c}{[km\,s$^{-1}$]} & filter \\
\noalign{\smallskip}\hline\noalign{\smallskip}
\multicolumn{1}{c}{(1)} &\multicolumn{1}{c}{(2)} &\multicolumn{1}{c}{(3)} &\multicolumn{1}{c}{(4)} &\multicolumn{1}{c}{(5)} &
\multicolumn{1}{c}{(6)} &\multicolumn{1}{c}{(7)} &\multicolumn{1}{c}{(8)} &\multicolumn{1}{c}{(9)} &\multicolumn{1}{c}{(10)} &
\multicolumn{1}{c}{(11)} \\
\noalign{\smallskip}\hline\noalign{\smallskip}
 1 & 30900 & 36260$^{\pm 260}$ & 81.1$^{\pm 0.1}$ & 0.324$^{\pm 0.002}$ & 0.74$^{\pm 0.01}$ & 0.42$^{\pm 0.01}$ & 0.358$^{\pm 0.005}$ & 73.5$^{\pm 0.3}$ & 183 & \\
\noalign{\smallskip}
 2 & 30410 & 29310$^{\pm 210}$ & 89.3$^{\pm 0.2}$ & 0.144$^{\pm 0.001}$ & 1.27$^{\pm 0.02}$ & 1.21$^{\pm 0.04}$ & 0.500$^{\pm 0.001}$ & 144.6$^{\pm 0.2}$ & 126 & \\
   & & 28740$^{+1810}_{-1170}$ & 89.8$^{+0.1}_{-1.6}$ & 0.144$^{+0.007}_{-0.004}$ & 1.30$^{+0.04}_{-0.19}$ & 1.29$^{+0.05}_{-0.30}$&&&&O \\
\noalign{\smallskip}
 3 & 35560 & 32510$^{\pm 200}$ & 77.1$^{\pm 0.1}$ & 0.445$^{\pm 0.003}$ & 0.53$^{\pm 0.01}$ & 0.57$^{\pm 0.01}$ & 0.036$^{\pm 0.004}$ & 283$^{\pm 2}$ & 240 & \\
\noalign{\smallskip}
 4 & 33040 & 29490$^{\pm 120}$ & 75.3$^{\pm 0.1}$ & 0.486$^{\pm 0.003}$ & \multicolumn{2}{c}{0.49$^{\pm 0.01}$} &0&...& 250 & \\
   & & 29220$^{+1560}_{-1740}$ & 75.0$^{+1.9}_{-1.3}$ & 0.485$^{+0.025}_{-0.030}$ & \multicolumn{2}{c}{0.48$^{+0.13}_{-0.03}$}&&&&MR \\
\noalign{\smallskip}
 5 & 34120 & 32760$^{\pm 90}$  & 77.5$^{\pm 0.1}$ & 0.461$^{\pm 0.002}$ & 0.53$^{\pm 0.01}$ & 0.54$^{\pm 0.01}$ & 0.042$^{\pm 0.001}$ & 337$^{\pm 3}$ & 229 & \\
   & & 32660$^{+1130}_{-670}$  & 77.3$^{+0.9}_{-0.5}$ & 0.463$^{+0.009}_{-0.013}$ & 0.53$^{+0.08}_{-0.03}$ & 0.54$^{+0.08}_{-0.03}$ &&&&O\\
\noalign{\smallskip}
 6 & 32430 & 25170$^{\pm 100}$ & 80.2$^{\pm 0.1}$ & 0.422$^{\pm 0.002}$ & 0.63$^{\pm 0.01}$ & 0.65$^{\pm 0.01}$ & 0.033$^{\pm 0.003}$ & 254$^{\pm 2}$ & 235 &\\
\noalign{\smallskip}
 7 & 32580 & 30880$^{\pm 150}$ & 84.9$^{\pm 0.1}$ & 0.341$^{\pm 0.002}$ & 0.77$^{\pm 0.01}$ & 0.93$^{\pm 0.01}$ & 0.271$^{\pm 0.005}$ & 260.9$^{\pm 0.2}$ & 219 & \\
\noalign{\smallskip}
 8 & 25060 & 18070$^{\pm 100}$ & 83.5$^{\pm 0.2}$ & 0.377$^{\pm 0.001}$ & 1.09$^{\pm 0.02}$ & 1.17$^{\pm 0.01}$ & 0.099$^{\pm 0.004}$ & 307$^{\pm 2}$ & 195 & \\
   & & 17600$^{+970}_{-930}$   & 84.1$^{+2.9}_{-1.8}$ & 0.368$^{+0.019}_{-0.019}$ & 1.11$^{+0.24}_{-0.13}$ & 1.18$^{+0.21}_{-0.12}$ &&&&O\\
\noalign{\smallskip}
 9 & 28120 & 26780$^{\pm 180}$ & 78.6$^{\pm 0.1}$ & 0.511$^{\pm 0.004}$ & \multicolumn{2}{c}{0.62$^{\pm 0.01}$} &0&...& 229 & \\
   & & 29080$^{+1170}_{-1080}$ & 78.8$^{+1.9}_{-0.5}$ & 0.512$^{+0.011}_{-0.025}$ & \multicolumn{2}{c}{0.63$^{+0.18}_{-0.03}$}&&&&O \\
\noalign{\smallskip}
10 & 29310 & 27560$^{\pm 100}$ & 71.5$^{\pm 0.1}$ & 0.595$^{\pm 0.003}$ & \multicolumn{2}{c}{0.47$^{\pm 0.01}$} &0&...& 262 & \\
   & & 28000$^{+690}_{-1270}$ & 71.5$^{+1.0}_{-0.7}$ & 0.595$^{+0.011}_{-0.016}$ & \multicolumn{2}{c}{0.48$^{+0.04}_{-0.03}$}&&&&O \\
\noalign{\smallskip}
11 & 26420 & 25090$^{\pm 120}$ & 89.3$^{\pm 1.2}$ & 0.510$^{\pm 0.001}$ & 1.15$^{\pm 0.05}$ & 1.15$^{\pm 0.05}$ & 0.014$^{\pm 0.002}$ & 220$^{\pm 10}$ & 194 & \\
   & & 24940$^{+830}_{-820}$ & 89.8$^{+0.2}_{-3.4}$ & 0.513$^{+0.013}_{-0.014}$ & 1.17$^{+0.03}_{-0.14}$ & 1.17$^{+0.03}_{-0.14}$&&&&O\\
\noalign{\smallskip}
12 & 28710 & 28340$^{\pm 160}$ & 70.9$^{\pm 0.1}$ & 0.584$^{\pm 0.006}$ & \multicolumn{2}{c}{0.44$^{\pm 0.01}$} &0&...& 265 & \\
   & & 27980$^{+1250}_{-1230}$ & 70.7$^{+1.3}_{-1.1}$ & 0.592$^{+0.013}_{-0.023}$ & \multicolumn{2}{c}{0.44$^{+0.06}_{-0.02}$}&&&&O \\
\noalign{\smallskip}
13 & 29380 & 28350$^{\pm 120}$ & 84.1$^{\pm 0.1}$ & 0.335$^{\pm 0.002}$ & 0.71$^{\pm 0.01}$ & 0.89$^{\pm 0.01}$ & 0.282$^{\pm 0.004}$ & 281.5$^{\pm 0.1}$ & 202 &\\
\noalign{\smallskip}
14 & 27420 & 26060$^{\pm 110}$ & 72.8$^{\pm 0.1}$ & 0.530$^{\pm 0.003}$ & \multicolumn{2}{c}{0.45$^{\pm 0.01}$} &0&...& 220 & \\
   & & 26020$^{+1140}_{-1700}$ & 73.1$^{+1.3}_{-1.3}$ & 0.527$^{+0.021}_{-0.024}$ & \multicolumn{2}{c}{0.46$^{+0.08}_{-0.03}$}&&&&MB\\
\noalign{\smallskip}
15 & 26980 & 26960$^{\pm 200}$ & 81.8$^{\pm 0.1}$ & 0.245$^{\pm 0.002}$ & \multicolumn{2}{c}{0.50$^{\pm 0.01}$} &0&...& 152 & \\
   & & 27060$^{+2080}_{-2900}$ & 81.5$^{+4.7}_{-1.1}$ & 0.258$^{+0.025}_{-0.051}$ & \multicolumn{2}{c}{0.42$^{+0.76}_{-0.04}$}&&&&MR\\
\noalign{\smallskip}
16 & 22860 & 20480$^{\pm 140}$ & 85.2$^{\pm 0.3}$ & 0.368$^{\pm 0.002}$ & 1.10$^{\pm 0.02}$ & 1.15$^{\pm 0.02}$ & 0.079$^{\pm 0.006}$ & 288$^{\pm 1}$ & 151 & \\
   & & 20230$^{+1160}_{-760}$ & 86.1$^{+3.9}_{-3.8}$ & 0.367$^{+0.031}_{-0.017}$ & 1.17$^{+0.30}_{-0.28}$ & 1.21$^{+0.27}_{-0.26}$&&&&O\\
\noalign{\smallskip}
17 & 26730 & 24340$^{\pm 80}$ & 79.2$^{\pm 0.1}$ & 0.502$^{\pm 0.002}$ & \multicolumn{2}{c}{0.64$^{\pm 0.01}$} &0&...& 221 & \\
   & & 24600$^{+780}_{-1140}$ & 79.4$^{+1.8}_{-0.4}$ & 0.503$^{+0.009}_{-0.021}$ & \multicolumn{2}{c}{0.93$^{+0.17}_{-0.02}$}&&&&O \\
\noalign{\smallskip}
18 & 25940 & 25270$^{\pm 180}$ & 83.6$^{\pm 0.1}$ & 0.241$^{\pm 0.001}$ & 0.70$^{\pm 0.01}$ & 0.53$^{\pm 0.01}$ & 0.180$^{\pm 0.005}$ & 116.2$^{\pm 0.9}$ & 139 & \\
   & & 25470$^{+1090}_{-5530}$ & 83.6$^{+2.8}_{-0.4}$ & 0.236$^{+0.009}_{-0.026}$ & 0.66$^{+0.42}_{-0.06}$ & 0.50$^{+0.45}_{-0.06}$&&&&O\\
\noalign{\smallskip}
19 & 26000 & 25590$^{\pm 80}$ & 86.9$^{\pm 0.1}$ & 0.467$^{\pm 0.003}$ & 0.91$^{\pm 0.01}$ & 0.90$^{\pm 0.01}$ & 0.078$^{\pm 0.002}$ & 149$^{\pm 2}$ & 194 & \\
   & & 25550$^{+760}_{-900}$ & 87.0$^{+3.0}_{-0.7}$ & 0.473$^{+0.013}_{-0.022}$ & 0.90$^{+0.20}_{-0.03}$ & 0.89$^{+0.21}_{-0.03}$&&&&O\\
\noalign{\smallskip}
20 & 26060 & 24510$^{\pm 130}$ & 84.4$^{\pm 0.1}$ & 0.307$^{\pm 0.002}$ & 0.65$^{\pm 0.01}$ & 0.78$^{\pm 0.01}$ & 0.206$^{\pm 0.004}$ & 278.7$^{\pm 0.2}$ & 165 & \\
   & & 25430$^{+2630}_{-1030}$ & 84.4$^{+1.9}_{-0.6}$ & 0.309$^{+0.016}_{-0.017}$ & 0.65$^{+0.27}_{-0.05}$ & 0.80$^{+0.24}_{-0.05}$&&&&O\\
\noalign{\smallskip}
21 & 27040 & 21340$^{\pm 100}$ & 81.9$^{\pm 0.1}$ & 0.355$^{\pm 0.002}$ & 0.58$^{\pm 0.01}$ & 0.73$^{\pm 0.01}$ & 0.215$^{\pm 0.004}$ & 294.8$^{\pm 0.4}$ & 192 &\\
\noalign{\smallskip}
22 & 26000 & 19280$^{\pm 190}$ & 73.8$^{\pm 0.1}$ & 0.417$^{\pm 0.004}$ & 0.38$^{\pm 0.01}$ & 0.32$^{\pm 0.01}$ & 0.060$^{\pm 0.003}$ & 128$^{\pm 3}$ & 195 & \\
   & & 16960$^{+1160}_{-1840}$ & 74.2$^{+1.6}_{-1.2}$ & 0.416$^{+0.021}_{-0.029}$ & 0.39$^{+0.08}_{-0.02}$ & 0.33$^{+0.07}_{-0.02}$&&&&O\\
\noalign{\smallskip}
23 & 25240 & 27010$^{\pm 130}$ & 72.4$^{\pm 0.1}$ & 0.510$^{\pm 0.004}$ & \multicolumn{2}{c}{0.41$^{\pm 0.01}$} &0&...& 237 & \\
   & & 23180$^{+1960}_{-2000}$ & 72.2$^{+2.5}_{-1.9}$ & 0.512$^{+0.032}_{-0.048}$ & \multicolumn{2}{c}{0.42$^{+0.13}_{-0.04}$}&&&&MB\\
\noalign{\smallskip}
24 & 27920 & 26550$^{\pm 150}$ & 86.7$^{\pm 0.2}$ & 0.361$^{\pm 0.001}$ & 1.08$^{\pm 0.02}$ & 1.15$^{\pm 0.01}$ & 0.157$^{\pm 0.005}$ & 284.8$^{\pm 0.4}$ & 191 &\\
\noalign{\smallskip}
25 & 25640 & 24730$^{\pm 280}$ & 83.8$^{\pm 0.1}$ & 0.171$^{\pm 0.002}$ & \multicolumn{2}{c}{0.40$^{\pm 0.01}$} &0&...& 131 & \\
   & & 24190$^{+2160}_{-2230}$ & 84.2$^{+2.3}_{-0.8}$ & 0.164$^{+0.017}_{-0.029}$ & \multicolumn{2}{c}{0.41$^{+0.59}_{-0.05}$}&&&&O\\
\noalign{\smallskip}\hline
\end{tabular} 
\end{table*}
\vfill\eject

\setcounter{table}{2}
\begin{table*}
\centering
\caption{Continued.}
\begin{tabular}{rclllccclcc}
\hline\noalign{\smallskip}
\#& $T_1^{*}$ & \multicolumn{1}{c}{$T_2$}& \multicolumn{1}{c}{$i$} & \multicolumn{1}{c}{$r_1 + r_2$} & 
   \multicolumn{1}{c}{$F_{\rm e,1}$} & \multicolumn{1}{c}{$F_{\rm e,2}$} & \multicolumn{1}{c}{$e$} & \multicolumn{1}{c}{$\omega$} &
   \multicolumn{1}{c}{$K_{1,2}$} & M-C \\
  &   [K]     & \multicolumn{1}{c}{[K]}  & \multicolumn{1}{c}{[$^{\rm o}$]} &&&&&\multicolumn{1}{c}{[$^{\rm o}$]} & 
   \multicolumn{1}{c}{[km\,s$^{-1}$]} & filter \\
\noalign{\smallskip}\hline\noalign{\smallskip}
26 & 26300 & 24780$^{\pm 100}$ & 84.7$^{\pm 0.2}$ & 0.485$^{\pm 0.001}$ & \multicolumn{2}{c}{1.00$^{\pm 0.01}$} &0&...& 216 & \\
   & & 25600$^{+1040}_{-800}$ & 85.1$^{+4.3}_{-3.8}$ & 0.483$^{+0.040}_{-0.024}$ & \multicolumn{2}{c}{1.02$^{+0.21}_{-0.31}$}&&&&MB\\
\noalign{\smallskip}
27 & 24380 & 25520$^{\pm 100}$ & 87.6$^{\pm 0.2}$ & 0.404$^{\pm 0.003}$ & 0.95$^{\pm 0.01}$ & 0.95$^{\pm 0.01}$ & 0.052$^{\pm 0.001}$ & 185$^{\pm 5}$ & 200 & \\
   & & 25360$^{+520}_{-560}$ & 88.8$^{+1.2}_{-0.7}$ & 0.403$^{+0.010}_{-0.012}$ & 0.98$^{+0.08}_{-0.05}$ & 0.98$^{+0.08}_{-0.05}$&&&&O\\
\noalign{\smallskip}
28 & 25180 & 22150$^{\pm 60}$ & 79.8$^{\pm 0.1}$ & 0.486$^{\pm 0.002}$ & \multicolumn{2}{c}{0.64$^{\pm 0.01}$} &0&...& 215 & \\
   & & 22350$^{+780}_{-1290}$ & 80.1$^{+6.1}_{-0.8}$ & 0.483$^{+0.017}_{-0.042}$ & \multicolumn{2}{c}{0.66$^{+0.44}_{-0.03}$}&&&&MB\\
\noalign{\smallskip}
29 & 24660 & 21940$^{\pm 50}$ & 86.9$^{\pm 0.2}$ & 0.568$^{\pm 0.001}$ & \multicolumn{2}{c}{1.05$^{\pm 0.01}$} &0&...& 252 & \\
   & & 21880$^{+1490}_{-580}$ & 87.0$^{+3.0}_{-4.9}$ & 0.570$^{+0.024}_{-0.018}$ & \multicolumn{2}{c}{1.05$^{+0.12}_{-0.27}$}&&&&O \\
\noalign{\smallskip}
30 & 26980 & 23600$^{\pm 150}$ & 79.0$^{\pm 0.1}$ & 0.330$^{\pm 0.003}$ & \multicolumn{2}{c}{0.43$^{\pm 0.01}$} &0&...& 194 & \\
   & & 23070$^{+1620}_{-1930}$ & 79.1$^{+3.3}_{-0.9}$ & 0.326$^{+0.018}_{-0.038}$ & \multicolumn{2}{c}{0.45$^{+0.38}_{-0.04}$}&&&&O \\
\noalign{\smallskip}
31 & 23990 & 21160$^{\pm 160}$ & 72.6$^{\pm 0.1}$ & 0.477$^{\pm 0.005}$ & \multicolumn{2}{c}{0.38$^{\pm 0.01}$} &0&...& 223 & \\
   & & 21630$^{+1880}_{-2350}$ & 72.7$^{+3.1}_{-2.0}$ & 0.479$^{+0.034}_{-0.056}$ & \multicolumn{2}{c}{0.38$^{+0.16}_{-0.03}$}&&&&MB\\
\noalign{\smallskip}
32 & 23770 & 22590$^{\pm 110}$ & 80.2$^{\pm 0.1}$ & 0.461$^{\pm 0.003}$ & \multicolumn{2}{c}{0.64$^{\pm 0.01}$} &0&...& 188 & \\
   & & 22650$^{+860}_{-990}$ & 76.7$^{+1.2}_{-0.9}$ & 0.462$^{+0.018}_{-0.021}$ & \multicolumn{2}{c}{0.52$^{+0.11}_{-0.03}$}&&&&MB\\
\noalign{\smallskip}
33 & 23990 & 20310$^{\pm 50}$ & 88.0$^{\pm 0.2}$ & 0.433$^{\pm 0.001}$ & 1.07$^{\pm 0.09}$ & 1.08$^{\pm 0.08}$ & 0.037$^{\pm 0.003}$ & 280.2$^{\pm 0.8}$ & 213& \\
\noalign{\smallskip}
34 & 23230 & 22520$^{\pm 120}$ & 72.5$^{\pm 0.1}$ & 0.455$^{\pm 0.003}$ & \multicolumn{2}{c}{0.34$^{\pm 0.01}$} &0&...& 198 & \\
   & & 22140$^{+2250}_{-1410}$ & 72.1$^{+2.1}_{-1.9}$ & 0.459$^{+0.033}_{-0.036}$ & \multicolumn{2}{c}{0.33$^{+0.10}_{-0.02}$}&&&&MB\\
\noalign{\smallskip}
35 & 22750 & 22130$^{\pm 140}$ & 89.5$^{\pm 2.0}$ & 0.411$^{\pm 0.002}$ & 1.20$^{\pm 0.10}$ & 1.19$^{\pm 0.10}$ & 0.134$^{\pm 0.001}$ & 178$^{\pm 2}$ & 179 & \\
   & & 22520$^{+760}_{-950}$ & 89.8$^{+0.2}_{-3.6}$ & 0.411$^{+0.015}_{-0.014}$ & 1.21$^{+0.04}_{-0.19}$ & 1.21$^{+0.04}_{-0.29}$&&&&MB\\
\noalign{\smallskip} 
36 & 22540 & 23360$^{\pm 400}$ & 87.7$^{\pm 0.2}$ & 0.153$^{\pm 0.001}$ & 0.92$^{\pm 0.04}$ & 1.06$^{\pm 0.03}$ & 0.252$^{\pm 0.006}$ & 234$^{\pm 1}$ & 106 & \\
   & & 22770$^{+5370}_{-4590}$ & 87.2$^{+2.8}_{-2.1}$ & 0.157$^{+0.038}_{-0.032}$ & 0.82$^{+0.65}_{-0.34}$ & 0.98$^{+0.50}_{-0.32}$&&&&MR\\
\noalign{\smallskip}
37 & 24660 & 19940$^{\pm 180}$ & 87.3$^{\pm 0.1}$ & 0.180$^{\pm 0.001}$ & 0.99$^{\pm 0.01}$ & 0.89$^{\pm 0.02}$ & 0.179$^{\pm 0.005}$ &  60.6$^{\pm 0.9}$ & 118 & \\
   & & 19120$^{+3300}_{-2420}$ & 87.9$^{+2.1}_{-1.7}$ & 0.170$^{+0.027}_{-0.022}$ & 1.06$^{+0.34}_{-0.34}$ & 0.98$^{+0.43}_{-0.36}$&&&&MR\\
\noalign{\smallskip}
38 & 23440 & 19330$^{\pm 40}$ & 83.6$^{\pm 0.1}$ & 0.587$^{\pm 0.002}$ & \multicolumn{2}{c}{0.93$^{\pm 0.01}$} &0&...& 232 & \\
   & & 19430$^{+550}_{-1180}$ & 83.8$^{+4.1}_{-1.5}$ & 0.588$^{+0.026}_{-0.021}$ & \multicolumn{2}{c}{0.93$^{+0.15}_{-0.14}$}&&&&MB\\
\noalign{\smallskip}
39 & 21280 & 19560$^{\pm 140}$ & 77.2$^{\pm 0.1}$ & 0.449$^{\pm 0.003}$ & 0.59$^{\pm 0.01}$ & 0.53$^{\pm 0.01}$ & 0.067$^{\pm 0.004}$ &  59$^{\pm 2}$ & 161 & \\
   & & 19160$^{+1990}_{-2490}$ & 77.8$^{+6.0}_{-1.5}$ & 0.447$^{+0.025}_{-0.063}$ & 0.58$^{+0.50}_{-0.06}$ & 0.53$^{+0.51}_{-0.06}$&&&&O \\
\noalign{\smallskip}
40 & 22230 & 20560$^{\pm 280}$ & 77.1$^{\pm 0.2}$ & 0.423$^{\pm 0.004}$ & 0.44$^{\pm 0.01}$ & 0.52$^{\pm 0.01}$ & 0.089$^{\pm 0.006}$ & 236$^{\pm 2}$ & 202 & \\
   & & 20420$^{+1600}_{-1520}$ & 77.2$^{+2.3}_{-0.9}$ & 0.421$^{+0.021}_{-0.038}$ & 0.44$^{+0.20}_{-0.03}$ & 0.52$^{+0.20}_{-0.03}$&&&&O \\
\noalign{\smallskip}
41 & 21230 & 21680$^{\pm 450}$ & 83.6$^{\pm 0.1}$ & 0.200$^{\pm 0.002}$ & 0.64$^{\pm 0.01}$ & 0.36$^{\pm 0.02}$ & 0.230$^{\pm 0.010}$ & 105.0$^{\pm 0.7}$ & 132 & \\
   & & 22220$^{+2280}_{-8440}$ & 84.0$^{+2.2}_{-1.1}$ & 0.194$^{+0.024}_{-0.025}$ & 0.66$^{+0.44}_{-0.10}$ & 0.39$^{+0.38}_{-0.09}$&&&&O \\
\noalign{\smallskip}
42 & 24270 & 22050$^{\pm 220}$ & 82.1$^{\pm 0.1}$ & 0.271$^{\pm 0.001}$ & 0.99$^{\pm 0.01}$ & 0.62$^{\pm 0.01}$ & 0.279$^{\pm 0.005}$ & 124.4$^{\pm 0.8}$ & 122 & \\
   & & 21410$^{+8260}_{-8460}$ & 82.3$^{+4.8}_{-4.0}$ & 0.264$^{+0.064}_{-0.027}$ & 1.03$^{+0.37}_{-0.49}$ & 0.65$^{+0.58}_{-0.39}$&&&&O \\
\noalign{\smallskip}
43 & 24210 & 21540$^{\pm 210}$ & 86.8$^{\pm 0.2}$ & 0.172$^{\pm 0.001}$ & 0.98$^{\pm 0.02}$ & 0.99$^{\pm 0.02}$ & 0.025$^{\pm 0.001}$ & 182$^{\pm 19}$ & 111 & \\
   & & 21150$^{+1950}_{-1250}$ & 87.5$^{+2.5}_{-2.1}$ & 0.165$^{+0.023}_{-0.017}$ & 1.07$^{+0.40}_{-0.49}$ & 1.07$^{+0.39}_{-0.49}$&&&&O \\
\noalign{\smallskip}
44 & 20800 & 21150$^{\pm 130}$ & 84.9$^{\pm 0.1}$ & 0.300$^{\pm 0.002}$ & 0.88$^{\pm 0.01}$ & 0.68$^{\pm 0.01}$ & 0.326$^{\pm 0.004}$ & 108.0$^{\pm 0.2}$ & 136 & \\
   & & 20750$^{+980}_{-3430}$ & 85.1$^{+2.4}_{-0.5}$ & 0.295$^{+0.011}_{-0.018}$ & 0.88$^{+0.22}_{-0.09}$ & 0.69$^{+0.30}_{-0.09}$&&&&O \\
\noalign{\smallskip}
45 & 23170 & 23130$^{\pm 160}$ & 81.4$^{\pm 0.1}$ & 0.372$^{\pm 0.002}$ & 0.82$^{\pm 0.01}$ & 0.74$^{\pm 0.01}$ & 0.089$^{\pm 0.004}$ &  60$^{\pm 2}$ & 193& \\
\noalign{\smallskip}
46 & 22030 & 21450$^{\pm 140}$ & 84.4$^{\pm 0.1}$ & 0.396$^{\pm 0.003}$ & 0.79$^{\pm 0.01}$ & 0.80$^{\pm 0.01}$ & 0.078$^{\pm 0.003}$ & 201$^{\pm 4}$ & 181 & \\
   & & 21420$^{+990}_{-1120}$ & 85.2$^{+2.7}_{-1.6}$ & 0.392$^{+0.024}_{-0.026}$ & 0.91$^{+0.20}_{-0.17}$ & 0.92$^{+0.20}_{-0.17}$&&&&O \\
\noalign{\smallskip}
47 & 21880 & 21140$^{\pm 110}$ & 86.0$^{\pm 0.1}$ & 0.225$^{\pm 0.002}$ & 0.77$^{\pm 0.01}$ & 0.70$^{\pm 0.01}$ & 0.113$^{\pm 0.005}$ &  74.2$^{\pm 0.7}$ & 146 & \\
   & & 21170$^{+1250}_{-1710}$ & 85.9$^{+2.3}_{-0.7}$ & 0.226$^{+0.014}_{-0.023}$ & 0.77$^{+0.34}_{-0.07}$ & 0.69$^{+0.38}_{-0.07}$&&&&MB\\
\noalign{\smallskip}
48 & 22590 & 16480$^{\pm 80}$ & 79.8$^{\pm 0.1}$ & 0.447$^{\pm 0.002}$ & \multicolumn{2}{c}{0.62$^{\pm 0.01}$} &0&...& 218 & \\
   & & 16740$^{+810}_{-1380}$ & 79.2$^{+4.0}_{-0.9}$ & 0.449$^{+0.021}_{-0.032}$ & \multicolumn{2}{c}{0.64$^{+0.29}_{-0.05}$}&&&&MB\\
\noalign{\smallskip}
49 & 21140 & 19680$^{\pm 50}$ & 82.0$^{\pm 0.1}$ & 0.546$^{\pm 0.001}$ & \multicolumn{2}{c}{0.75$^{\pm 0.01}$} &0&...& 233 & \\
   & & 19310$^{+1160}_{-670}$ & 81.7$^{+3.6}_{-0.7}$ & 0.548$^{+0.016}_{-0.034}$ & \multicolumn{2}{c}{0.74$^{+0.27}_{-0.02}$}&&&&MR\\
\noalign{\smallskip}
50 & 21140 & 19030$^{\pm 130}$ & 83.0$^{\pm 0.1}$ & 0.259$^{\pm 0.001}$ & 0.60$^{\pm 0.01}$ & 0.65$^{\pm 0.01}$ & 0.090$^{\pm 0.002}$ & 330$^{\pm 2}$ & 123 & \\
   & & 19750$^{+1150}_{-1600}$ & 83.4$^{+2.9}_{-1.3}$ & 0.257$^{+0.023}_{-0.033}$ & 0.68$^{+0.37}_{-0.20}$ & 0.73$^{+0.36}_{-0.21}$&&&&MB\\
\noalign{\smallskip}
51 & 21630 & 21570$^{\pm 70}$ & 78.9$^{\pm 0.1}$ & 0.524$^{\pm 0.002}$ & \multicolumn{2}{c}{0.64$^{\pm 0.01}$} &0&...& 187 & \\
   & & 21750$^{+1100}_{-1230}$ & 78.8$^{+2.5}_{-1.0}$ & 0.525$^{+0.017}_{-0.036}$ & \multicolumn{2}{c}{0.64$^{+0.22}_{-0.02}$}&&&&MB\\
\noalign{\smallskip}\hline
\end{tabular} 
\end{table*}
\vfill\eject

\setcounter{table}{2}
\begin{table*}
\centering
\caption{Continued.}
\begin{tabular}{rclllccclcc}
\hline\noalign{\smallskip}
\#& $T_1^{*}$ & \multicolumn{1}{c}{$T_2$}& \multicolumn{1}{c}{$i$} & \multicolumn{1}{c}{$r_1 + r_2$} & 
   \multicolumn{1}{c}{$F_{\rm e,1}$} & \multicolumn{1}{c}{$F_{\rm e,2}$} & \multicolumn{1}{c}{$e$} & \multicolumn{1}{c}{$\omega$} &
   \multicolumn{1}{c}{$K_{1,2}$} & M-C \\
  &   [K]     & \multicolumn{1}{c}{[K]}  & \multicolumn{1}{c}{[$^{\rm o}$]} &&&&&\multicolumn{1}{c}{[$^{\rm o}$]} & 
   \multicolumn{1}{c}{[km\,s$^{-1}$]} & filter \\
\noalign{\smallskip}\hline\noalign{\smallskip}
52 & 21090 & 22670$^{\pm 220}$ & 78.9$^{\pm 0.1}$ & 0.376$^{\pm 0.002}$ & 0.56$^{\pm 0.01}$ & 0.46$^{\pm 0.01}$ & 0.108$^{\pm 0.005}$ & 114$^{\pm 1}$ & 183 &\\
\noalign{\smallskip}
53 & 21780 & 20180$^{\pm 120}$ & 72.8$^{\pm 0.1}$ & 0.571$^{\pm 0.004}$ & \multicolumn{2}{c}{0.49$^{\pm 0.01}$} &0&...& 249 & \\
   & & 20390$^{+1020}_{-1290}$ & 72.9$^{+1.7}_{-1.4}$ & 0.568$^{+0.024}_{-0.031}$ & \multicolumn{2}{c}{0.49$^{+0.09}_{-0.03}$}&&&&MB\\
\noalign{\smallskip}
54 & 21380 & 20310$^{\pm 130}$ & 82.1$^{\pm 0.2}$ & 0.570$^{\pm 0.005}$ & \multicolumn{2}{c}{0.78$^{\pm 0.01}$} &0&...& 193 & \\
   & & 20460$^{+900}_{-1300}$ & 82.4$^{+3.4}_{-1.1}$ & 0.565$^{+0.021}_{-0.034}$ & \multicolumn{2}{c}{0.78$^{+0.25}_{-0.03}$}&&&&O\\
\noalign{\smallskip}
55 & 21330 & 21230$^{\pm 130}$ & 78.2$^{\pm 0.1}$ & 0.372$^{\pm 0.003}$ & \multicolumn{2}{c}{0.45$^{\pm 0.01}$} &0&...& 167 & \\
   & & 21540$^{+2460}_{-2180}$ & 78.1$^{+5.8}_{-1.4}$ & 0.374$^{+0.031}_{-0.072}$ & \multicolumn{2}{c}{0.47$^{+0.61}_{-0.05}$}&&&&MR\\
\noalign{\smallskip}
56 & 20750 & 20500$^{\pm 60}$ & 85.1$^{\pm 0.1}$ & 0.632$^{\pm 0.003}$ & \multicolumn{2}{c}{0.88$^{\pm 0.01}$} &0&...& 232 & \\
   & & 20620$^{+720}_{-1040}$ & 85.5$^{+4.5}_{-1.2}$ & 0.633$^{+0.015}_{-0.026}$ & \multicolumn{2}{c}{0.91$^{+0.21}_{-0.06}$}&&&&MB\\
\noalign{\smallskip}
57 & 19680 & 19340$^{\pm 100}$ & 87.9$^{\pm 0.4}$ & 0.490$^{\pm 0.002}$ & \multicolumn{2}{c}{1.03$^{\pm 0.02}$} &0&...& 153 & \\
   & & 19330$^{+770}_{-600}$ & 88.3$^{+1.7}_{-2.4}$ & 0.486$^{+0.024}_{-0.015}$ & \multicolumn{2}{c}{1.05$^{+0.08}_{-0.16}$}&&&&O \\
\noalign{\smallskip}
58 & 22860 & 19940$^{\pm 130}$ & 89.3$^{\pm 0.7}$ & 0.271$^{\pm 0.001}$ & 1.43$^{\pm 0.07}$ & 1.45$^{\pm 0.06}$ & 0.214$^{\pm 0.003}$ & 320.0$^{\pm 0.9}$ & 156 & \\
   & & 19960$^{+1530}_{-1610}$ & 89.3$^{+0.7}_{-3.0}$ & 0.271$^{+0.019}_{-0.013}$ & 1.43$^{+0.12}_{-0.32}$ & 1.45$^{+0.10}_{-0.25}$&&&&O \\
\noalign{\smallskip}
59 & 21040 & 19060$^{\pm 110}$ & 81.8$^{\pm 0.1}$ & 0.390$^{\pm 0.003}$ & 0.68$^{\pm 0.01}$ & 0.64$^{\pm 0.01}$ & 0.053$^{\pm 0.005}$ & 106$^{\pm 2}$ & 177 &\\
\noalign{\smallskip}
60 & 19060 & 18360$^{\pm 90}$ & 85.5$^{\pm 0.1}$ & 0.280$^{\pm 0.002}$ & 0.75$^{\pm 0.01}$ & 0.75$^{\pm 0.01}$ & 0.243$^{\pm 0.001}$ & 180$^{\pm 1}$ & 139 & \\
   & & 18400$^{+1050}_{-910}$ & 85.7$^{+4.3}_{-0.8}$ & 0.282$^{+0.018}_{-0.029}$ & 0.77$^{+0.46}_{-0.05}$ & 0.77$^{+0.46}_{-0.05}$&&&&O \\
\noalign{\smallskip}
61 & 20370 & 18490$^{\pm 90}$ & 84.6$^{\pm 0.1}$ & 0.260$^{\pm 0.001}$ & 0.65$^{\pm 0.01}$ & 0.68$^{\pm 0.01}$ & 0.230$^{\pm 0.001}$ & 345$^{\pm 1}$ & 160 & \\
   & & 18730$^{+1490}_{-1100}$ & 84.9$^{+2.5}_{-0.7}$ & 0.255$^{+0.017}_{-0.023}$ & 0.69$^{+0.35}_{-0.06}$ & 0.73$^{+0.34}_{-0.06}$&&&&O \\
\noalign{\smallskip}
62 & 20990 & 18020$^{\pm 200}$ & 83.1$^{\pm 0.1}$ & 0.335$^{\pm 0.004}$ & 0.58$^{\pm 0.02}$ & 0.76$^{\pm 0.01}$ & 0.296$^{\pm 0.007}$ & 248.3$^{\pm 0.6}$ & 161 &\\
\noalign{\smallskip}
63 & 20510 & 19260$^{\pm 150}$ & 87.4$^{\pm 0.1}$ & 0.267$^{\pm 0.001}$ & 0.81$^{\pm 0.01}$ & 0.93$^{\pm 0.01}$ & 0.365$^{\pm 0.005}$ & 264.5$^{\pm 0.1}$ & 156 & \\
\noalign{\smallskip}
64 & 20650 & 19330$^{\pm 250}$ & 87.3$^{\pm 0.1}$ & 0.139$^{\pm 0.001}$ & 0.97$^{\pm 0.02}$ & 0.90$^{\pm 0.02}$ & 0.068$^{\pm 0.010}$ &  77$^{\pm 3}$ &  96 & \\
   & & 19380$^{+3120}_{-1960}$ & 87.4$^{+2.6}_{-2.1}$ & 0.138$^{+0.029}_{-0.023}$ & 1.00$^{+0.52}_{-0.46}$ & 0.93$^{+0.59}_{-0.46}$&&&&O \\
\noalign{\smallskip}
65 & 20230 & 19690$^{\pm 90}$ & 85.7$^{\pm 0.1}$ & 0.475$^{\pm 0.004}$ & \multicolumn{2}{c}{0.85$^{\pm 0.01}$} &0&...& 202 & \\
   & & 19660$^{+820}_{-920}$ & 86.1$^{+3.9}_{-1.2}$ & 0.476$^{+0.016}_{-0.026}$ & \multicolumn{2}{c}{0.88$^{+0.25}_{-0.05}$}&&&&MB\\
\noalign{\smallskip}
66 & 19820 & 19050$^{\pm 70}$ & 84.1$^{\pm 0.1}$ & 0.414$^{\pm 0.002}$ & 0.76$^{\pm 0.01}$ & 0.76$^{\pm 0.01}$ & 0.108$^{\pm 0.001}$ &   7$^{\pm 2}$ & 175 & \\
   & & 19290$^{+610}_{-860}$ & 84.4$^{+2.1}_{-0.6}$ & 0.411$^{+0.016}_{-0.018}$ & 0.79$^{+0.22}_{-0.04}$ & 0.78$^{+0.22}_{-0.04}$&&&&O \\
\noalign{\smallskip}
67 & 21330 & 22700$^{\pm 200}$ & 79.7$^{\pm 0.1}$ & 0.415$^{\pm 0.004}$ & 0.69$^{\pm 0.01}$ & 0.59$^{\pm 0.01}$ & 0.106$^{\pm 0.007}$ &  74$^{\pm 1}$ & 184 & \\
   & & 21360$^{+2100}_{-1810}$ & 79.6$^{+5.7}_{-1.5}$ & 0.414$^{+0.035}_{-0.061}$ & 0.61$^{+0.55}_{-0.03}$ & 0.52$^{+0.58}_{-0.03}$&&&&O \\
\noalign{\smallskip}
68 & 20650 & 14180$^{\pm 90}$ & 88.0$^{\pm 0.5}$ & 0.384$^{\pm 0.001}$ & \multicolumn{2}{c}{1.31$^{\pm 0.03}$} &0&...& 204 & \\
   & & 14180$^{+1790}_{-1180}$ & 88.5$^{+1.5}_{-11.9}$ & 0.385$^{+0.096}_{-0.019}$ & \multicolumn{2}{c}{1.34$^{+0.13}_{-0.82}$}&&&&MB\\
\noalign{\smallskip}
69 & 20840 & 20800$^{\pm 190}$ & 79.9$^{\pm 0.1}$ & 0.298$^{\pm 0.003}$ & \multicolumn{2}{c}{0.42$^{\pm 0.01}$} &0&...& 135 & \\
   & & 20210$^{+3910}_{-2480}$ & 81.4$^{+8.6}_{-2.8}$ & 0.275$^{+0.051}_{-0.073}$ & \multicolumn{2}{c}{0.60$^{+1.27}_{-0.21}$}&&&&O \\
\noalign{\smallskip}
70 & 19770 & 19860$^{\pm 100}$ & 86.6$^{\pm 0.1}$ & 0.253$^{\pm 0.001}$ & 0.78$^{\pm 0.01}$ & 0.79$^{\pm 0.01}$ & 0.017$^{\pm 0.003}$ & 218$^{\pm 11}$ & 116 & \\
   & & 19960$^{+1050}_{-1110}$ & 85.9$^{+2.7}_{-0.8}$ & 0.256$^{+0.018}_{-0.026}$ & 0.78$^{+0.37}_{-0.09}$ & 0.78$^{+0.37}_{-0.09}$&&&&MB\\
\noalign{\smallskip}
71 & 20140 & 18350$^{\pm 130}$ & 83.6$^{\pm 0.2}$ & 0.349$^{\pm 0.002}$ & \multicolumn{2}{c}{0.96$^{\pm 0.02}$} &0&...& 151 & \\
   & & 18420$^{+1800}_{-1700}$ & 83.1$^{+6.9}_{-3.6}$ & 0.355$^{+0.056}_{-0.043}$ & \multicolumn{2}{c}{0.92$^{+0.51}_{-0.35}$}&&&&MB\\
\noalign{\smallskip}
72 & 20000 & 20240$^{\pm 90}$ & 82.0$^{\pm 0.1}$ & 0.500$^{\pm 0.003}$ & \multicolumn{2}{c}{0.74$^{\pm 0.01}$} &0&...& 188 & \\
   & & 20250$^{+760}_{-1140}$ & 82.0$^{+3.1}_{-0.7}$ & 0.501$^{+0.019}_{-0.030}$ & \multicolumn{2}{c}{0.73$^{+0.27}_{-0.02}$}&&&&O \\
\noalign{\smallskip}
73 & 21700 & 20900$^{\pm 180}$ & 89.4$^{\pm 0.8}$ & 0.320$^{\pm 0.002}$ & 1.05$^{\pm 0.05}$ & 1.05$^{\pm 0.04}$ & 0.324$^{\pm 0.002}$ & 341$^{\pm 1}$ & 151 & \\
   & & 20900$^{+1610}_{-1560}$ & 89.9$^{+0.1}_{-2.6}$ & 0.318$^{+0.021}_{-0.021}$ & 1.09$^{+0.07}_{-0.21}$ & 1.10$^{+0.07}_{-0.19}$&&&&MR\\
\noalign{\smallskip}
74 & 20140 & 18850$^{\pm 270}$ & 83.3$^{\pm 0.1}$ & 0.228$^{\pm 0.002}$ & 0.57$^{\pm 0.01}$ & 0.56$^{\pm 0.01}$ & 0.021$^{\pm 0.004}$ &  30$^{\pm 14}$ & 115 & \\
   & & 18790$^{+3420}_{-2810}$ & 83.3$^{+6.6}_{-2.1}$ & 0.224$^{+0.038}_{-0.065}$ & 0.62$^{+1.10}_{-0.19}$ & 0.60$^{+1.12}_{-0.18}$&&&&MR\\
\noalign{\smallskip}
75 & 18920 & 18850$^{\pm 50}$ & 88.8$^{\pm 0.3}$ & 0.530$^{\pm 0.001}$ & \multicolumn{2}{c}{1.02$^{\pm 0.01}$} &0&...& 189 & \\
   & & 18790$^{+790}_{-960}$ & 89.5$^{+0.5}_{-3.3}$ & 0.530$^{+0.025}_{-0.018}$ & \multicolumn{2}{c}{1.05$^{+0.05}_{-0.17}$}&&&&MR\\
\noalign{\smallskip}
76 & 17950 & 17690$^{\pm 110}$ & 87.8$^{\pm 0.3}$ & 0.298$^{\pm 0.001}$ & 1.16$^{\pm 0.02}$ & 1.17$^{\pm 0.02}$ & 0.268$^{\pm 0.001}$ & 353$^{\pm 1}$ & 137 & \\
   & & 17680$^{+1070}_{-1020}$ & 89.7$^{+0.3}_{-4.6}$ & 0.291$^{+0.028}_{-0.013}$ & 1.30$^{+0.06}_{-0.35}$ & 1.30$^{+0.06}_{-0.33}$&&&&MB\\
\noalign{\smallskip}
77 & 21230 & 18480$^{\pm 100}$ & 86.2$^{\pm 0.3}$ & 0.408$^{\pm 0.001}$ & 1.14$^{\pm 0.02}$ & 1.12$^{\pm 0.02}$ & 0.051$^{\pm 0.004}$ & 127$^{\pm 3}$ & 196 & \\
   & & 18560$^{+1060}_{-1030}$ & 86.3$^{+3.6}_{-2.1}$ & 0.403$^{+0.020}_{-0.024}$ & 1.15$^{+0.22}_{-0.13}$ & 1.13$^{+0.24}_{-0.14}$&&&&O \\
\noalign{\smallskip}\hline
\end{tabular} 
\end{table*}

\vfill\eject

\setcounter{table}{2}
\begin{table*}
\centering
\caption{Continued.}
\begin{tabular}{rclllccclcc}
\hline\noalign{\smallskip}
\#& $T_1^{*}$ & \multicolumn{1}{c}{$T_2$}& \multicolumn{1}{c}{$i$} & \multicolumn{1}{c}{$r_1 + r_2$} & 
   \multicolumn{1}{c}{$F_{\rm e,1}$} & \multicolumn{1}{c}{$F_{\rm e,2}$} & \multicolumn{1}{c}{$e$} & \multicolumn{1}{c}{$\omega$} &
   \multicolumn{1}{c}{$K_{1,2}$} & M-C \\
  &   [K]     & \multicolumn{1}{c}{[K]}  & \multicolumn{1}{c}{[$^{\rm o}$]} &&&&&\multicolumn{1}{c}{[$^{\rm o}$]} & 
   \multicolumn{1}{c}{[km\,s$^{-1}$]} & filter \\
\noalign{\smallskip}\hline\noalign{\smallskip}
78 & 19770 & 19600$^{\pm 110}$ & 79.0$^{\pm 0.1}$ & 0.434$^{\pm 0.003}$ & 0.56$^{\pm 0.01}$ & 0.58$^{\pm 0.01}$ & 0.048$^{\pm 0.002}$ & 211$^{\pm 5}$ & 178 & \\
   & & 20050$^{+1290}_{-1330}$ & 79.1$^{+3.9}_{-1.2}$ & 0.437$^{+0.027}_{-0.045}$ & 0.56$^{+0.41}_{-0.02}$ & 0.58$^{+0.41}_{-0.02}$&&&&MB\\
\noalign{\smallskip}
79 & 20320 & 20240$^{\pm 110}$ & 77.3$^{\pm 0.2}$ & 0.541$^{\pm 0.003}$ & \multicolumn{2}{c}{0.68$^{\pm 0.01}$} &0&...& 192 & \\
   & & 20450$^{+960}_{-2080}$ & 77.5$^{+6.7}_{-2.0}$ & 0.542$^{+0.037}_{-0.073}$ & \multicolumn{2}{c}{0.69$^{+0.44}_{-0.12}$}&&&&MB\\
\noalign{\smallskip}
80 & 19770 & 16910$^{\pm 90}$ & 79.4$^{\pm 0.1}$ & 0.490$^{\pm 0.003}$ & \multicolumn{2}{c}{0.66$^{\pm 0.01}$} &0&...& 203 & \\
   & & 16850$^{+750}_{-1320}$ & 79.4$^{+10.0}_{-0.9}$ & 0.488$^{+0.021}_{-0.065}$ & \multicolumn{2}{c}{0.67$^{+0.63}_{-0.06}$}&&&&MB\\
\noalign{\smallskip}
81 & 19360 & 14440$^{\pm 170}$ & 86.6$^{\pm 0.2}$ & 0.237$^{\pm 0.002}$ & 1.01$^{\pm 0.03}$ & 1.12$^{\pm 0.02}$ & 0.448$^{\pm 0.003}$ & 333.0$^{\pm 0.8}$ & 162 & \\
   & & 13930$^{+1600}_{-1520}$ & 84.7$^{+2.6}_{-1.2}$ & 0.258$^{+0.020}_{-0.025}$ & 0.77$^{+0.24}_{-0.17}$ & 0.92$^{+0.20}_{-0.18}$&&&&O \\
\noalign{\smallskip}
82 & 15920 & 16010$^{\pm 170}$ & 83.6$^{\pm 0.1}$ & 0.282$^{\pm 0.003}$ & 0.64$^{\pm 0.01}$ & 0.63$^{\pm 0.01}$ & 0.015$^{\pm 0.005}$ &  36$^{\pm 14}$ & 123 & \\
   & & 15840$^{+1780}_{-1890}$ & 83.6$^{+6.4}_{-1.9}$ & 0.288$^{+0.036}_{-0.052}$ & 0.71$^{+0.70}_{-0.15}$ & 0.70$^{+0.71}_{-0.15}$&&&&MR\\
\noalign{\smallskip}
83 & 19770 & 17920$^{\pm 90}$ & 88.4$^{\pm 0.4}$ & 0.493$^{\pm 0.002}$ & \multicolumn{2}{c}{1.03$^{\pm 0.02}$} &0&...& 223 & \\
   & & 18170$^{+570}_{-790}$ & 89.4$^{+0.6}_{-2.6}$ & 0.493$^{+0.020}_{-0.017}$ & \multicolumn{2}{c}{1.05$^{+0.05}_{-0.16}$}&&&&O \\
\noalign{\smallskip}
84 & 17860 & 16270$^{\pm 210}$ & 86.7$^{\pm 0.2}$ & 0.303$^{\pm 0.002}$ & 0.90$^{\pm 0.02}$ & 1.03$^{\pm 0.01}$ & 0.306$^{\pm 0.007}$ & 254.8$^{\pm 0.4}$ & 143 & \\
   & & 17100$^{+2440}_{-3370}$ & 87.6$^{+2.4}_{-3.2}$ & 0.302$^{+0.041}_{-0.034}$ & 0.99$^{+0.26}_{-0.36}$ & 1.09$^{+0.17}_{-0.29}$&&&&MB\\
\noalign{\smallskip}
85 & 19190 & 18630$^{\pm 220}$ & 77.8$^{\pm 0.1}$ & 0.402$^{\pm 0.004}$ & 0.51$^{\pm 0.01}$ & 0.44$^{\pm 0.01}$ & 0.120$^{\pm 0.004}$ & 145$^{\pm 3}$ & 183 & \\
   & & 17430$^{+1480}_{-1870}$ & 78.0$^{+5.0}_{-1.1}$ & 0.399$^{+0.026}_{-0.059}$ & 0.55$^{+0.49}_{-0.05}$ & 0.47$^{+0.50}_{-0.05}$&&&&O \\
\noalign{\smallskip}
86 & 19060 & 15840$^{\pm 250}$ & 85.5$^{\pm 0.2}$ & 0.258$^{\pm 0.002}$ & 0.93$^{\pm 0.02}$ & 0.81$^{\pm 0.02}$ & 0.234$^{\pm 0.006}$ & 135$^{\pm 2}$ & 159 & \\
   & & 16480$^{+1760}_{-2480}$ & 85.1$^{+3.8}_{-0.9}$ & 0.261$^{+0.019}_{-0.032}$ & 0.75$^{+0.41}_{-0.04}$ & 0.63$^{+0.49}_{-0.05}$&&&&O \\
\noalign{\smallskip}
87 & 17990 & 17560$^{\pm 60}$ & 82.7$^{\pm 0.2}$ & 0.569$^{\pm 0.003}$ & \multicolumn{2}{c}{0.84$^{\pm 0.01}$} &0&...& 194 & \\
   & & 17450$^{+830}_{-970}$ & 82.5$^{+4.2}_{-1.0}$ & 0.576$^{+0.018}_{-0.039}$ & \multicolumn{2}{c}{0.81$^{+0.26}_{-0.06}$}&&&&MB\\
\noalign{\smallskip}
88 & 19280 & 16780$^{\pm 50}$ & 89.3$^{\pm 1.1}$ & 0.541$^{\pm 0.001}$ & \multicolumn{2}{c}{1.18$^{\pm 0.04}$} &0&...& 216 & \\
   & & 16750$^{+580}_{-470}$ & 87.6$^{+2.4}_{-2.5}$ & 0.540$^{+0.015}_{-0.014}$ & \multicolumn{2}{c}{1.13$^{+0.12}_{-0.11}$}&&&&MB\\
\noalign{\smallskip}
89 & 20230 & 17760$^{\pm 130}$ & 80.2$^{\pm 0.1}$ & 0.414$^{\pm 0.002}$ & 0.64$^{\pm 0.01}$ & 0.60$^{\pm 0.01}$ & 0.052$^{\pm 0.005}$ & 71$^{\pm 2}$ & 201 &\\
\noalign{\smallskip}
90 & 17060 & 16900$^{\pm 380}$ & 87.8$^{\pm 0.2}$ & 0.226$^{\pm 0.002}$ & 0.89$^{\pm 0.03}$ & 1.04$^{\pm 0.02}$ & 0.406$^{\pm 0.006}$ & 262.1$^{\pm 0.1}$ & 131 & \\
   & & 16000$^{+4740}_{-2250}$ & 87.2$^{+2.8}_{-1.3}$ & 0.227$^{+0.034}_{-0.032}$ & 0.75$^{+0.51}_{-0.13}$ & 0.93$^{+0.33}_{-0.09}$&&&&MB\\
\noalign{\smallskip}
91 & 17950 & 14230$^{\pm 120}$ & 81.6$^{\pm 0.1}$ & 0.290$^{\pm 0.002}$ & \multicolumn{2}{c}{0.57$^{\pm 0.01}$} &0&...& 145 & \\
   & & 13790$^{+1830}_{-2240}$ & 81.4$^{+8.6}_{-1.6}$ & 0.295$^{+0.037}_{-0.081}$ & \multicolumn{2}{c}{0.58$^{+1.09}_{-0.11}$}&&&&MB\\
\noalign{\smallskip}
92 & 17950 & 17950$^{\pm 140}$ & 87.1$^{\pm 0.1}$ & 0.221$^{\pm 0.002}$ & 0.87$^{\pm 0.01}$ & 0.96$^{\pm 0.01}$ & 0.309$^{\pm 0.003}$ & 215.9$^{\pm 0.8}$ & 131 & \\
   & & 17990$^{+2240}_{-2240}$ & 87.5$^{+2.5}_{-1.7}$ & 0.218$^{+0.026}_{-0.086}$ & 0.93$^{+0.35}_{-0.27}$ & 1.01$^{+0.28}_{-0.24}$&&&&MB\\
\noalign{\smallskip}
93 & 18160 & 17580$^{\pm 90}$ & 83.3$^{\pm 0.1}$ & 0.414$^{\pm 0.003}$ & \multicolumn{2}{c}{0.77$^{\pm 0.01}$} &0&...& 139 & \\
   & & 17340$^{+1220}_{-960}$ & 83.4$^{+3.5}_{-1.2}$ & 0.416$^{+0.023}_{-0.036}$ & \multicolumn{2}{c}{0.75$^{+0.32}_{-0.06}$}&&&&MR\\
\noalign{\smallskip}
94 & 18200 & 17550$^{\pm 130}$ & 85.9$^{\pm 0.2}$ & 0.283$^{\pm 0.002}$ & \multicolumn{2}{c}{1.00$^{\pm 0.02}$} &0&...& 138 & \\
   & & 17480$^{+1720}_{-1440}$ & 86.0$^{+4.0}_{-3.4}$ & 0.281$^{+0.052}_{-0.033}$ & \multicolumn{2}{c}{0.99$^{+0.40}_{-0.37}$}&&&&MB\\
\noalign{\smallskip}
95 & 16440 & 16530$^{\pm 90}$ & 86.2$^{\pm 0.2}$ & 0.356$^{\pm 0.002}$ & 0.99$^{\pm 0.01}$ & 0.95$^{\pm 0.01}$ & 0.211$^{\pm 0.002}$ & 150$^{\pm 1}$ & 132 & \\
   & & 16420$^{+1160}_{-760}$ & 86.2$^{+3.8}_{-2.0}$ & 0.356$^{+0.026}_{-0.021}$ & 0.98$^{+0.23}_{-0.21}$ & 0.93$^{+0.28}_{-0.22}$&&&&MB\\
\noalign{\smallskip}
96 & 18540 & 17900$^{\pm 100}$ & 81.9$^{\pm 0.1}$ & 0.421$^{\pm 0.002}$ & 0.78$^{\pm 0.01}$ & 0.75$^{\pm 0.01}$ & 0.103$^{\pm 0.003}$ & 155$^{\pm 2}$ & 170 & \\
   & & 17690$^{+1490}_{-1550}$ & 81.8$^{+8.2}_{-1.6}$ & 0.425$^{+0.034}_{-0.055}$ & 0.74$^{+0.58}_{-0.09}$ & 0.71$^{+0.62}_{-0.09}$&&&&MB\\
\noalign{\smallskip}
97 & 16440 & 15600$^{\pm 120}$ & 78.7$^{\pm 0.1}$ & 0.432$^{\pm 0.003}$ & 0.58$^{\pm 0.01}$ & 0.57$^{\pm 0.01}$ & 0.019$^{\pm 0.002}$ & 166$^{\pm 15}$ & 167 & \\
   & & 15700$^{+1160}_{-1810}$ & 80.8$^{+4.0}_{-2.7}$ & 0.415$^{+0.047}_{-0.039}$ & 0.75$^{+0.32}_{-0.19}$ & 0.74$^{+0.32}_{-0.19}$&&&&O \\
\noalign{\smallskip}
98 & 17180 & 16930$^{\pm 120}$ & 86.5$^{\pm 0.3}$ & 0.304$^{\pm 0.002}$ & \multicolumn{2}{c}{1.04$^{\pm 0.02}$} &0&...& 143 & \\
   & & 16950$^{+1140}_{-1150}$ & 86.5$^{+3.4}_{-3.5}$ & 0.304$^{+0.038}_{-0.030}$ & \multicolumn{2}{c}{1.04$^{+0.27}_{-0.40}$}&&&&MB\\
\noalign{\smallskip}\hline
\end{tabular} 
\end{table*}

\end{document}